%% file: paper.tex
\documentclass[acmtog,nonacm]{acmart}

\usepackage{booktabs} % For formal tables

\usepackage[ruled]{algorithm2e} % For algorithms

\SetAlFnt{\small}
\SetAlCapFnt{\small}
\SetAlCapNameFnt{\small}
\SetAlCapHSkip{0pt}

\makeatletter
\def\@ACM@copyright@check@cc{}
\makeatother
\setcopyright{cc}
\setcctype{by}
\acmJournal{TOG}
\acmYear{2025} 
\acmVolume{44} 
\acmNumber{4} 
\acmArticle{} 
\acmMonth{8} 
\acmPrice{}
\acmDOI{10.1145/3730866}

\usepackage{enumitem}
\usepackage{multirow}
\usepackage{wrapfig}

\newcommand{\piecewisesurf}{\mathcal{R}_{\text{p}}}
\newcommand{\rulcoefone}{a}
\newcommand{\rulcoeftwo}{b}
\newcommand{\facerule}{\mathbf{r}}
\newcommand{\facepointrule}{\overline{\mathbf{r}}}
\newcommand{\facept}{\mathbf{s}}
\newcommand{\facevtx}[2]{\mathbf{v}_{#1,#2}}
\newcommand{\faceedge}[2]{\mathbf{h}_{#1,#2}}

\newcommand{\pdir}{\mathbf{b}}
\newcommand{\pcurv}{\kappa}
\newcommand{\normalcurv}{\kappa_{n}}
\newcommand{\sffterm}{E_{\text{curv}}}

\newcommand{\closeterm}{E_{\text{close}}}

\newcommand{\displapterm}{E_{\text{lap}}}
\newcommand{\edgelengthterm}{E_{\text{len}}}

\newcommand{\welschterm}{E_{\text{sparse}}}
\newcommand{\barrierterm}{E_{\text{barr}}}
\newcommand{\edgelengthweight}{w_4}
\newcommand{\lapweight}{w_3}
\newcommand{\closeweight}{w_1}
\newcommand{\barrierweight}{w_2}

\newcommand{\initlocalcoord}{\mathbf{y}}
\newcommand{\smoothterm}{E_{\text{smo}}}
\newcommand{\initalignterm}{E_{\text{align}}}
\newcommand{\proxyinitalignterm}{\widehat{E}_{\text{align}}}
\newcommand{\unitterm}{E_{\text{unit}}}
\newcommand{\surrogunitterm}{\overline{E}_{\text{unit}}}
\newcommand{\surroginitalignterm}{\overline{E}_{\text{align}}}
\newcommand{\unitweight}{\mu_1}
\newcommand{\npfacealignweight}{\mu_2}
\newcommand{\pfacealignweight}{\mu_3}

\newcommand{\tangentdir}{\mathbf{d}}
\newcommand{\tangentplane}[1]{T_{#1}}
\newcommand{\sfftensor}{\mathbf{M}}
\newcommand{\centroid}{\mathbf{o}}
\newcommand{\sfffunc}[2]{E({#1},#2)}
\newcommand{\geoddispvec}[2]{\mathbf{g}_{#1,#2}}
\newcommand{\segmentdir}{\mathbf{s}}
\newcommand{\edgegeoderrfunc}{E_{\text{geod}}}
\newcommand{\welshthreshold}{\epsilon_{\text{edge}}}
\newcommand{\candidateedgeset}{\mathcal{C}_{\text{B}}}
\newcommand{\boundaryfaceset}{\mathcal{F}_{\text{B}}}
\newcommand{\facecomponent}{\mathcal{S}}
\newcommand{\combedgeerr}{E_{\text{comb}}}
\newcommand{\facecompedgeset}{\mathcal{E}_{\mathcal{S}}}
\newcommand{\facecompintedgeset}{\mathcal{E}_{\text{I}}^{\mathcal{S}}}
\newcommand{\edgecutcost}{E_{\text{cut}}}
\newcommand{\facelabelcost}{E_{\text{label}}}
\newcommand{\unlabelledfaceset}{\widetilde{\mathcal{F}}_{\facecomponent}}
\newcommand{\boundarysmoothterm}{E_{\text{bdr}}}

\newcommand{\rulingsmoothterm}{E_{\text{rul}}}

\newcommand{\finalalignterm}{E_{\text{disp}}}

\newcommand{\rulingsampleset}{\mathbf{S}_{\text{r}}}
\newcommand{\bdrsampleset}{\mathbf{S}_{\text{b}}}
\newcommand{\alignptweight}{\alpha}
\newcommand{\patchsmoothptweight}{\beta}
\newcommand{\bdrsmoothptweight}{\eta}
\newcommand{\rulingsamplept}{\mathbf{s}}
\newcommand{\bdrsamplept}{\mathbf{b}}
\newcommand{\normalcurvfunc}{K_\text{r}}
\newcommand{\bdrcurvfunc}{K_{\text{b}}}
\newcommand{\bdrsmoothtermweight}{\lambda_4}
\newcommand{\rulingsmoothtermweight}{\lambda_3}
\newcommand{\finalalignintweight}{\lambda_1}
\newcommand{\finalalignbdrweight}{\lambda_2}
\newcommand{\faceorthobases}{\mathbf{D}}

\newcommand{\pdirbases}{\mathbf{B}}
\newcommand{\asympdirset}{\mathcal{A}}
\newcommand{\curvaturecoef}{\mu}
\DeclareMathOperator*{\argmin}{arg\,min}
\newcommand{\facelabelcostweight}{\lambda_{\text{label}}}
\newcommand{\surfparampt}{\mathbf{p}}
\newcommand{\surfparamcurv}{\mathbf{a}}
\newcommand{\surfparamdir}{\mathbf{r}}
\newcommand{\surfparambdrcurv}{\mathbf{b}}
\newcommand{\numcompbdrsegs}{N}
\newcommand{\singlesplitvtx}{\mathbf{q}}
\newcommand{\virtualsplitvtx}{\mathbf{q}^{\ast}}
\newcommand{\smallareathreshold}{\epsilon_{A}}
\newcommand{\maxerr}{\epsilon_{\max}}
\newcommand{\avgerr}{\epsilon_{\textrm{avg}}}
\newcommand{\seamlength}{L_{\text{s}}}
\newcommand{\edgecurlerrfunc}{E_{\text{curl}}}

\begin{document}
% Title portion
\title{Piecewise Ruled Approximation for Freeform Mesh Surfaces}

% DO NOT ENTER AUTHOR INFORMATION FOR ANONYMOUS TECHNICAL PAPER SUBMISSIONS TO SIGGRAPH 2019!
\author{Yiling Pan}
\orcid{0000-0001-7076-3335}
%\affiliation{%
%  \institution{Tsinghua University}
%  \city{Beijing}
\email{pyl16@mails.tsinghua.edu.cn}
\author{Zhixin Xu}
\orcid{0009-0008-9486-7719}
%\affiliation{%
%  \institution{Tsinghua University}
%  \city{Beijing}
\email{xuzx23@mails.tsinghua.edu.cn}
\author{Bin Wang}
\authornote{Corresponding author.}
\orcid{0000-0002-5176-9202}
\email{wangbins@tsinghua.edu.cn}
\affiliation{%
  \institution{Tsinghua University}
  \city{Beijing}
  \country{China}}
\author{Bailin Deng}
\orcid{0000-0002-0158-7670}
\email{DengB3@cardiff.ac.uk}
\affiliation{%
  \institution{Cardiff University}
  \city{Cardiff}
  \country{United Kingdom}}

\renewcommand\shortauthors{Pan, Y. et al}
\begin{teaserfigure}
    \centering
    \includegraphics[width=\linewidth]{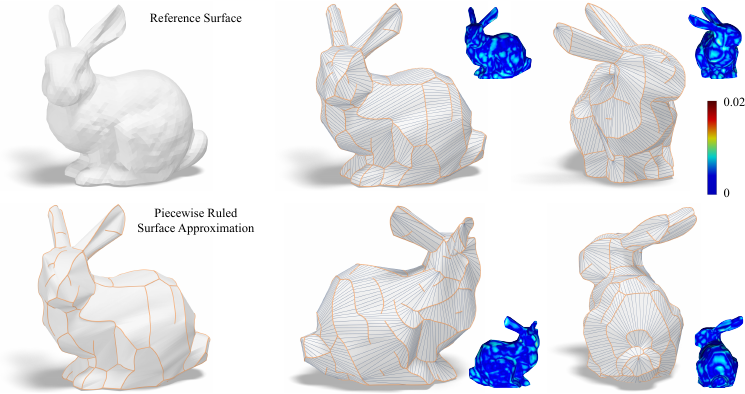}
    \caption{Piecewise ruled surface approximation computed using our method. The sub-figures on the right show different views of the result surface together with the rulings. We also use color coding to visualize the distance from the result surface to the reference surface, normalized by the reference surface's bounding box diagonal length.}
    \label{fig:teaser}
\end{teaserfigure}

\begin{abstract}
A ruled surface is a shape swept out by moving a line in 3D space. Due to their simple geometric forms, ruled surfaces have applications in various domains such as architecture and engineering. In the past, various approaches have been proposed to approximate a target shape using developable surfaces, which are special ruled surfaces with zero Gaussian curvature. However, methods for shape approximation using general ruled surfaces remain limited and often require the target shape to be either represented as parametric surfaces or have non-positive Gaussian curvature. In this paper, we propose a method to compute a piecewise ruled surface that approximates an arbitrary freeform mesh surface.  We first use a group-sparsity formulation to optimize the given mesh shape into an approximately piecewise ruled form, in conjunction with a tangent vector field that indicates the ruling directions. Afterward, we utilize the optimization result to extract seams that separate smooth families of rulings, and use the seams to construct the initial rulings. Finally, we further optimize the positions and orientations of the rulings to improve the alignment with the input target shape. We apply our method to a variety of freeform shapes with different topologies and complexity, demonstrating its effectiveness in approximating arbitrary shapes.
\end{abstract}
\maketitle





\input{introduction}

\input{related}
\input{method}
\input{experiment}

\input{conclusion}

\bibliographystyle{ACM-Reference-Format}
\bibliography{references}

\appendix

\input{supp}

\end{document}

%% file: introduction.tex
\section{Introduction}

\begin{figure}[t]
    \centering
    \includegraphics[height=0.35\linewidth]{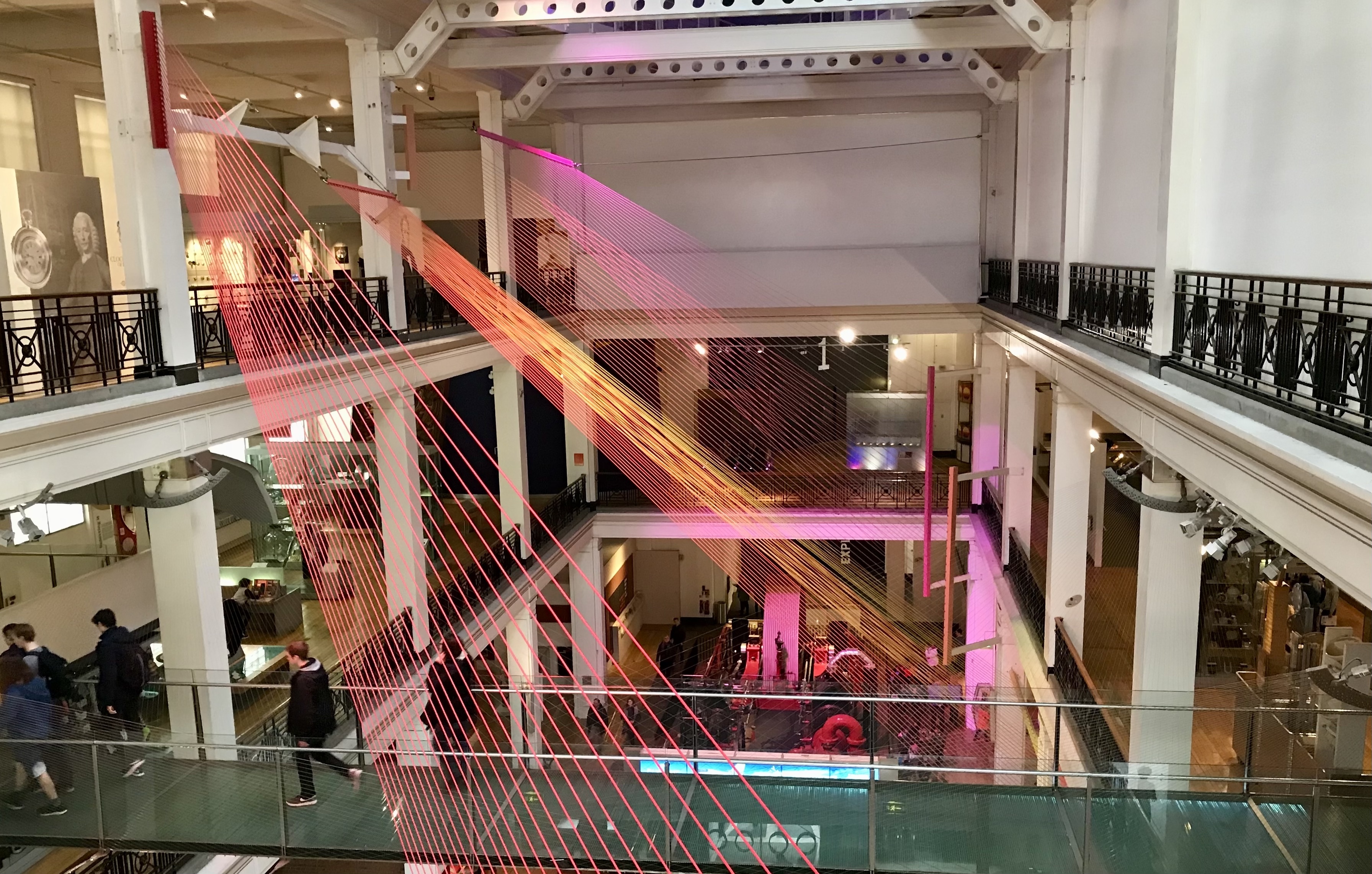}\hfill
    \includegraphics[height=0.35\linewidth]{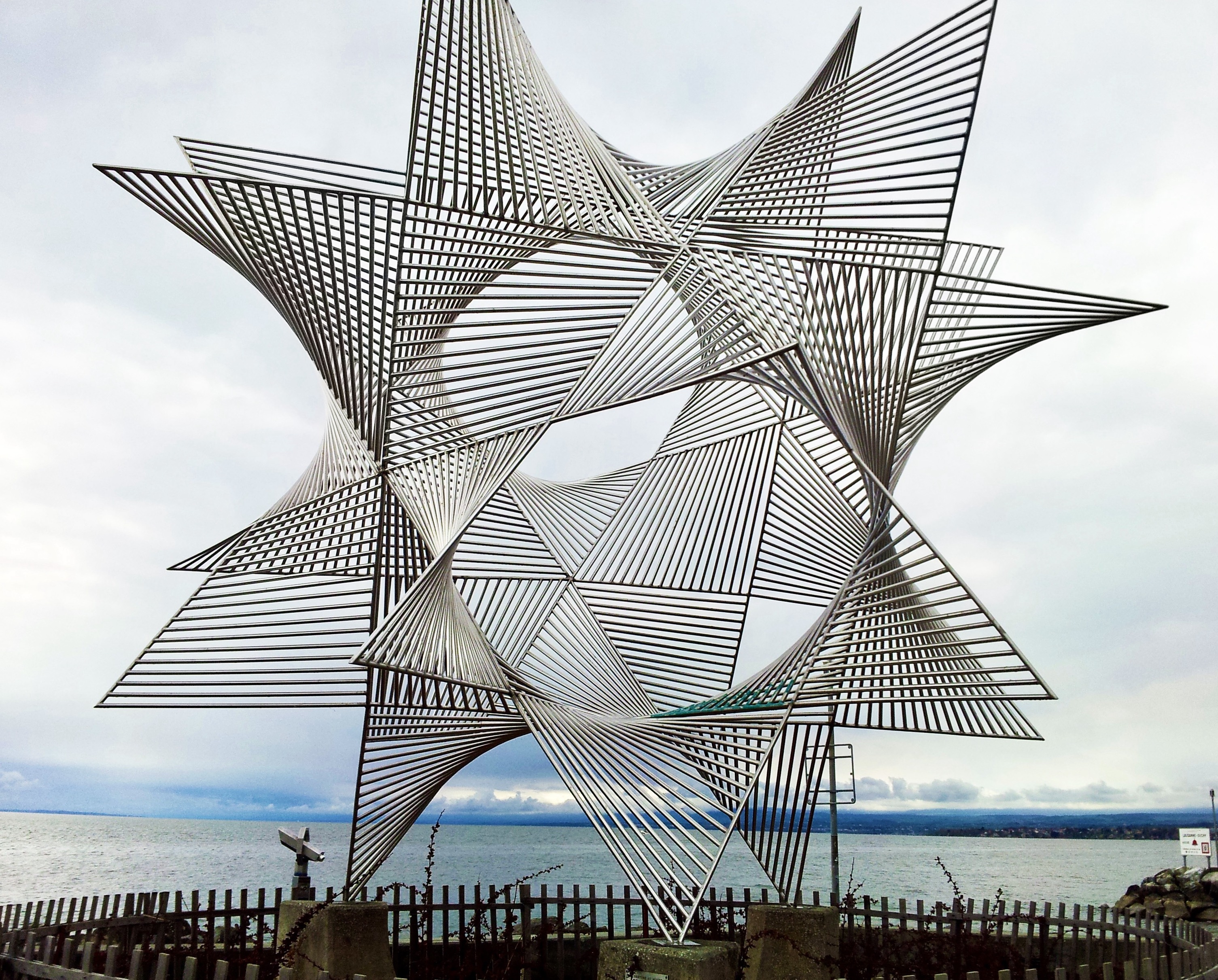}
    \caption{Examples of ruled surfaces. Left: an art installation using tensioned strings, in the Science Museum in London. Right: the `Ouverture au Monde' sculpture in Ouchy, Lausanne, by \'{A}ngel Duarte.}
    \label{fig:RuledSurfaceExamples}
\end{figure}

Freeform surfaces are widely used in various application domains, including computer-aided design, architecture, and manufacturing. However, their fabrication often poses significant challenges due to their complex geometry. One common approach to simplify the realization of such surfaces is to approximate them with simple elements. In this paper, we investigate freeform surface approximation using piecewise ruled surfaces.
A ruled surface is swept out by a continuous family of straight lines in 3D space. It can be parameterized by a correspondence between two curves $\surfparambdrcurv_1(u)$ and $\surfparambdrcurv_2(u)$:
\begin{equation}
    \surfparampt(u,v) =  (1-v) \surfparambdrcurv_1(u) + v \surfparambdrcurv_2(u),
\end{equation}
where points with the same $u$ value are on a straight line connecting the two curves.
Alternatively, a ruled surface can be represented as moving a straight line along a curve $\surfparamcurv(u)$:
\begin{equation}
    \surfparampt(u,v) =  \surfparamcurv(u) + v \surfparamdir(u),
\end{equation}
where $\surfparamdir(u)$ is the line direction at point $\surfparamcurv(u)$. The constituent straight lines on a ruled surface are called rulings.
The simple geometric form of ruled surfaces enables them to represent freeform shapes while allowing for efficient fabrication, e.g., by arranging straight elements such as tensioned strings (see Fig.~\ref{fig:RuledSurfaceExamples}) or using straight line motions of CNC tools like 5-axis side milling.

In the past, numerous research efforts have been devoted to piecewise approximation using developable patches, a special type of ruled surface with zero Gaussian curvature~\cite{Kilian2008,Tang2016,Stein2018,Ion2020,Zhao2022,Zhao2023}. However, investigation into approximation using general ruled surfaces\textemdash{}which allow for non-positive Gaussian curvature and offer more degrees of freedom than developable surfaces for better approximation\textemdash{}remains limited.
Existing methods have primarily focused on target surfaces represented in parametric forms~\cite{Elber1997,wang2014multi} or require the target surface to have non-positive Gaussian curvature everywhere~\cite{flory2013ruled}. This limitation restricts their applicability to a broader range of freeform surfaces that arise in real-world applications, such as those represented as triangle meshes and containing positive Gaussian curvature regions.
Moreover, for a complicated target shape where a single ruled surface patch is insufficient for good approximation, it is often necessary to introduce seams that separate adjacent smooth families of rulings to enable better alignment with the target, resulting in a piecewise ruled approximation (see Fig.~\ref{fig:teaser}).
Since such seams correspond to discontinuities of rulings, in practical applications it is preferable to reduce their presence for both aesthetics and ease of fabrication. On the other hand, an insufficient number of seams may lead to poor approximation due to limited degrees of freedom. It is a challenging problem to determine a suitable seam topology that can achieve good alignment with the target shape with as few seams as possible.
In this paper, we present a novel approach for computing a piecewise ruled surface that closely approximates an arbitrary freeform surface. Our method works on triangle meshes and can handle surfaces with both positive and negative Gaussian curvature, making it more versatile than previous techniques. 

We address this problem in three stages. First, we introduce a new approach that simultaneously optimizes the mesh shape and a ruling direction field on the mesh. Instead of the piecewise constant representation commonly used in geometry processing for tangent vector fields, we propose a novel representation of the ruling direction field based on its first-order approximation, which better captures the local variation of ruling directions. By employing a sparsity-based optimization with this representation, our method produces a piecewise smooth direction field on the updated mesh surface, which serves as an approximation of the final piecewise ruled surface shape and its ruling directions. The resulting direction field is smooth across most parts on the surface, with a small number of discontinuities that indicate potential seams.
Afterward, we utilize these discontinuities to determine the seam topology, and trace along the direction field to derive initial rulings with endpoints lying on the seams or the mesh boundary, resulting in an initial piecewise ruled surface. Lastly, we further optimize the seams and the surface boundary to control the rulings and improve their alignment with the target shape while maintaining the smoothness of the resulting shape. The final output is an explicit representation of rulings joining along the seams, ready for downstream applications.

Extensive experiments demonstrate that our method can produce piecewise ruled surfaces that accurately approximate various target freeform shapes with a limited presence of seams, outperforming existing approaches in terms of both accuracy and versatility.
In summary, the main contributions of this paper include:
\begin{itemize}[leftmargin=*,nosep]
\item A novel representation of ruling direction fields on mesh surfaces, which better captures their local variations than the conventional piecewise constant representation and enables more effective optimization of the ruling direction.
\item A sparsity-based formulation for the joint optimization of the mesh shape and the ruling direction field on the mesh, providing an effective approximation of the final piecewise ruled surface while reducing the presence of seams.
\item A topology extraction method that utilizes the optimized direction field to determine the layout of seams and initialize the ruling directions, facilitating further optimization to align the piecewise ruled surface with the target shape.
\end{itemize}

%% file: related.tex
\section{Related Work}

Ruled surface approximation and developable surface approximation have been extensively studied in computer graphics and computer-aided design. We review the most relevant papers in the following.

\paragraph{Ruled approximation}
Elber and Fish~\shortcite{Elber1997} recursively subdivide polynomial or rational tensor-product B-spline surfaces into patches that can be approximated using ruled surfaces until the approximation error is below a threshold. However, their method is not applicable to meshes, and the recursive subdivision may result in a large number of small patches.
Chen and Pottmann~\shortcite{chen1999approximation} approximate a given surface or scattered points with a ruled surface in tensor-product B-spline representation, by first constructing a sequence of lines approximating the target surface and then computing a smooth ruled surface according to the line sequence. Their approach can only produce a single ruled surface and cannot be used for piecewise ruled approximation.
Han et al.~\shortcite{Han2001} propose an isophote-based piecewise ruled surface approximation method for 5-axis machining, which incrementally searches for isophote regions with limited angles between the surface normals and a reference vector, and constructs rulings that connect cubic spline approximant curves for the boundaries of each isophote region. The incremental approach may produce a large number of patches; moreover, the method is not fully automatic and requires user interaction to handle isophote regions with complicated topology. 
Flöry et al.~\shortcite{flory2013ruled} approximate architectural freeform surfaces using smooth unions of ruled surface strips. Their method first initializes a set of ruling strips with the rulings aligned with asymptotic lines of the target surface, then optimizes the strips to align with the target shape while reducing the gaps between them and maintaining their shape smoothness. 
Their initialization requires an asymptotic direction field on the target surface, which only exists on surfaces with non-positive Gaussian curvature; therefore, the method is not directly applicable to surfaces with regions of positive Gaussian curvature.
Wang and Elber~\shortcite{wang2014multi} propose a multi-dimensional dynamic programming method to fit ruled surfaces to freeform rational surfaces by minimizing a discrete approximation error, which requires heavy sampling and GPU acceleration. Their method is not suitable for general triangle meshes and depends on structured parametric domains. In contrast, our method directly operates on meshes and better handles surfaces with complex shapes.
Steenstrup et al.~\shortcite{steenstrup2016cuttable} approximate a given surface with a piecewise ruled surface that is cuttable by wire cutting, which is then further processed with CNC milling to obtain the final shape. Their method requires the piecewise ruled surface to lie entirely on one side of the target surface and have no intersection with the target. While this is necessary to enable the subsequent CNC milling, it prevents the resulting piecewise ruled surface from achieving the best approximation for complex target shapes.
Hua and Jia~\shortcite{hua2018wire} perform piecewise ruled surface approximation of minimal surfaces for double-sided wire cutting, with the rulings aligned with a principal direction. Their method requires an analytic expression of the target minimal surface and hence lacks generalization to arbitrary freeform shapes.

Different from the above approaches, our method works directly on target surfaces represented as triangle meshes. Moreover, unlike~\cite{flory2013ruled} and \cite{hua2018wire}, our method allows arbitrary target surfaces with no restriction on their curvature. This makes our method more versatile than existing approaches.

\paragraph{Developable approximation}
Developable surfaces, characterized by their ability to be flattened onto a plane without stretching or tearing, are fundamental in manufacturing and design due to their ease of fabrication from sheet materials. There is extensive research in modeling and approximating shapes with developable surfaces.

Several approaches focus on creating discrete representations that capture developability. Mitani and Suzuki~\shortcite{Mitani2004} propose segmenting a mesh into feature-based parts and simplifying each part into unfoldable continuous triangle strips suitable for papercraft. Liu et al.~\shortcite{Liu2006} introduce conical meshes, a type of planar quad mesh aligned with principal curvatures, which is effective for architectural glass structures and offers a discrete model for developable surfaces. Rabinovich et al.~\shortcite{Rabinovich2018} model developables using discrete orthogonal geodesic nets based on angle constraints, avoiding explicit rulings but potentially allowing crumpled shapes without further regularization. Jiang et al.~\shortcite{Jiang2020} use quad-mesh-based isometric mappings via checkerboard patterns, offering flexibility for various modeling operations like cutting and folding, but distinct from our ruling-based piecewise approximation. Verhoeven et al.~\shortcite{Verhoeven2022} convert developable triangle meshes into planar quad strips by optimizing a scalar field whose isolines approximate rulings, differing from our direct optimization of rulings on the mesh itself. Unlike these methods that define specific discrete structures or rely on particular parameterizations, our approach directly optimizes a triangle mesh and a ruling field towards a piecewise ruled structure, handling arbitrary topologies and curvatures without requiring specific mesh types like quad meshes.

Another category involves deforming a given mesh towards developability. Stein et al.~\shortcite{Stein2018} define developability for triangle meshes based on the distribution of adjacent face normals for each vertex, and use a variational approach to drive a mesh toward developable patches. While this identifies seams implicitly, our method explicitly optimizes for rulings and seams based on the optimized field discontinuities. Binninger et al.~\shortcite{Binninger2021} iteratively thin the Gauss image of the input mesh while deforming the surface, preserving connectivity but differing from our joint optimization of shape and ruling field. Zhao et al.~\shortcite{Zhao2022} introduce an edge-oriented developability measure and optimize a deformation energy, followed by partitioning the deformed mesh. Later, Zhao et al.~\shortcite{Zhao2023} employ an evolutionary genetic algorithm to optimize a partition based on approximation error, patch count, and boundary length. Sell\'{a}n et al.~\shortcite{Sellan2020} focus on heightfields, casting developability as a rank minimization problem on the Hessian matrix, leading to a convex optimization that encourages piecewise developability. While these deformation-based methods produce (nearly) developable surfaces, they differ significantly from our approach, which targets the broader class of piecewise ruled surfaces and uses a distinct optimization strategy based on group sparsity and ruling field representation.

Interactive modeling and reconstruction methods have also been explored. Kilian et al.~\shortcite{Kilian2008} present an optimization framework for designing surfaces producible by curved folding, focusing on reconstructing surfaces from physical models or scans and maintaining global developability. Tang et al.~\shortcite{Tang2016} use splines to represent developables and formulate developability and curved folds as quadratic equations, enabling interactive modeling via an energy-guided projection solver. Ion et al.~\shortcite{Ion2020} approximate shapes by wrapping them with discrete orthogonal geodesic nets, using global optimization for patch placement followed by non-linear projection. These methods focus on specific design scenarios or different approximation strategies compared to our automatic piecewise ruled surface approximation from arbitrary meshes.

\begin{figure*}[!t]
        \centering
    \includegraphics[width=\linewidth]{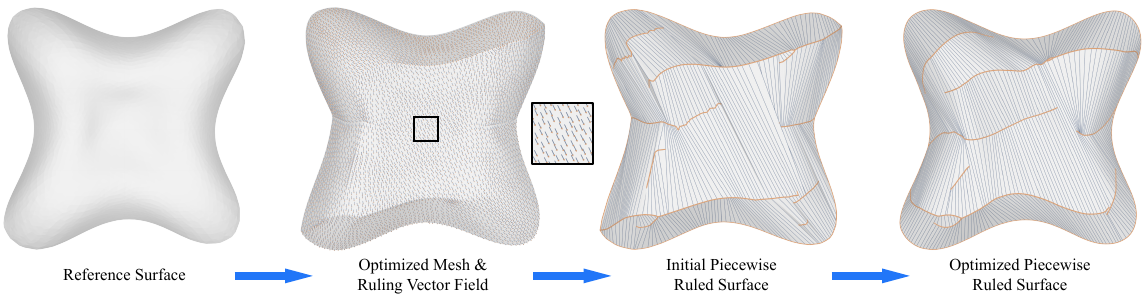}
    \caption{The pipeline of our piecewise ruled surface approximation. Given an input reference surface mesh, we first optimize the mesh shape together with a vector field on it, to approximate a piecewise ruled surface and its rulings. The result is then used to extract an initial piecewise ruled surface. Lastly, we optimize the piecewise ruled surface to further align it with the reference surface and obtain the final result.}
    \label{fig:pipeline}
\end{figure*}

%% file: method.tex
\section{Method}

Given a triangle mesh $\mathcal{M}$ representing a target freeform surface, we approximate its shape using a piecewise ruled surface $\piecewisesurf$; i.e., $\piecewisesurf$ may consist of multiple smooth families of rulings separated by a set of seams.
Similar to previous work on piecewise surface approximation~\cite{Stein2018,Binninger2021}, we refer to the ruling families separated by the seams as patches of the piecewise ruled surface, even though not all patches are of disk topology. In this way, the seams and boundaries of $\piecewisesurf$ become the boundaries of the patches.
Our method for piecewise ruled approximation consists of three steps (see Fig.~\ref{fig:pipeline}):
\begin{itemize}[leftmargin=*]
\item First, we optimize the vertex positions of $\mathcal{M}$ to deform it into an approximately piecewise ruled surface close to the target shape. A dense vector field on the mesh surface is also jointly optimized to indicate the ruling directions.
\item  Afterward, we use the optimized mesh and the vector field to determine a set of seams that separate adjacent families of rulings, and trace along the vector field to construct initial rulings for $\piecewisesurf$, with endpoints lying on the seams or the surface boundary. 
\item  Lastly, we optimize the seams and boundaries of $\piecewisesurf$ to control the rulings and further align them with the target shape, obtaining the final shape of $\piecewisesurf$.
\end{itemize}
Details for each step are explained in the following.

\subsection{Optimizing Mesh Shape and Ruling Direction Field}
Since a ruled surface has limited degrees of freedom and can only have non-positive Gaussian curvature, a complex target shape often requires multiple ruled patches to achieve a good approximation.
Choosing a suitable topology for such a piecewise ruled surface (i.e., the number of seams and their connectivity) is crucial for effectively optimizing its shape. However, the appropriate topology depends on the target shape in a highly nonlinear manner and cannot be determined easily. 
To address this problem, we first optimize the vertex positions of the input mesh $\mathcal{M}$ together with a vector field on its surface, such that $\mathcal{M}$ deforms into a shape consisting of multiple approximately ruled regions, where the vector field indicates the ruling directions.
The deformed mesh is then used for determining the piecewise ruled surface topology.
In this subsection, we explain our formulation for mesh optimization and its numerical solution. 
\begin{figure}[t]
    \centering
    \includegraphics[width=\linewidth]{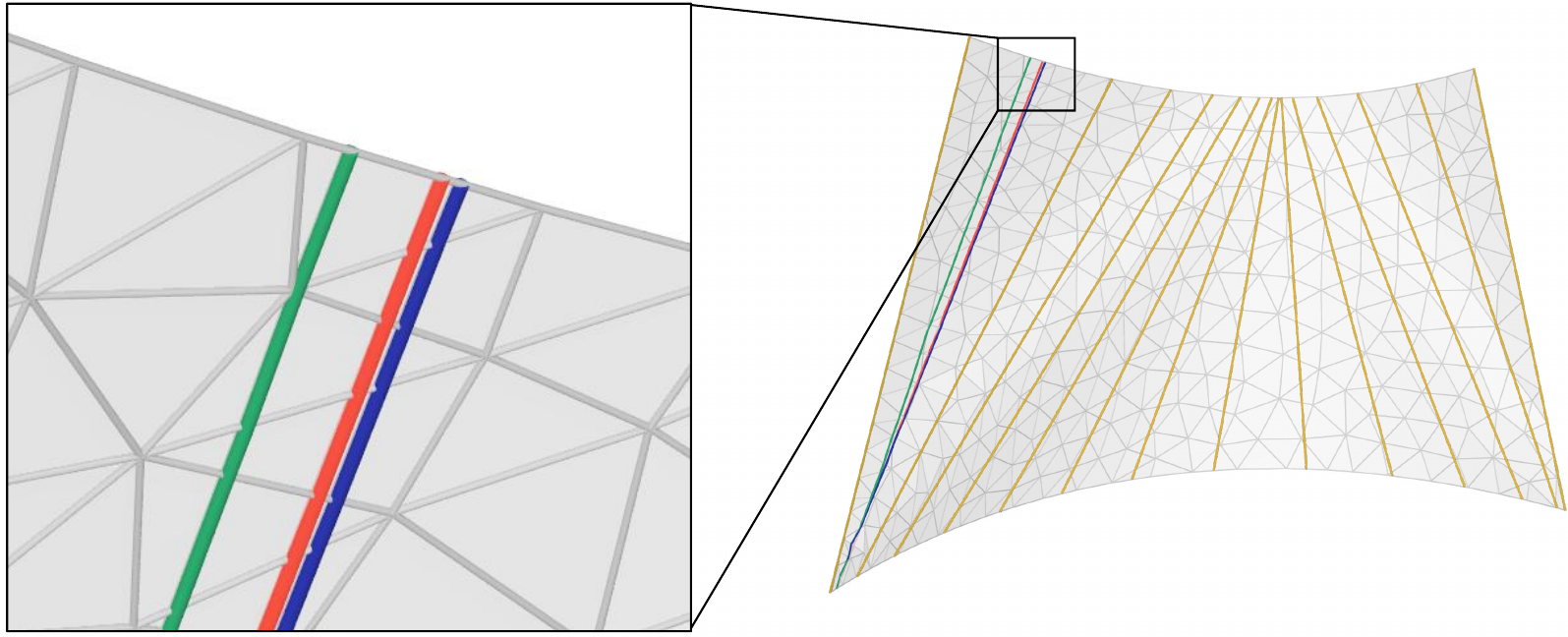}
    \caption{Piecewise constant vector fields are insufficient for capturing the variation of ruling directions on a ruled surface. Here we show a ruled surface discretized as a triangle mesh, where the original rulings are a smooth family of straight lines connecting the top and bottom curves (a subset is displayed here in brown). We first represent the ruling directions using a piecewise constant vector field according to the original ruling directions at the face centroids, and use it to trace a polyline starting from a point on the bottom boundary. There is a notable deviation between the resulting polyline (displayed in green) and the original ruling starting at the same point (displayed in red) when they reach the top boundary. On the other hand, if we represent the ruling directions with our vector field model in Eq.~\eqref{eq:PointRuilngVector} and start tracing from the same point, the resulting polyline (displayed in blue) is much closer to the original ruling.}
    \label{fig:rep-rfileds}
\end{figure}

\subsubsection{Representation of ruling direction fields}
On a smooth ruled surface, the direction vector of a ruling is a member of the tangent space at every point on the ruling.
Thus, we can represent the ruling directions as a tangent vector field on the mesh surface. The conventional approach is to use a piecewise-constant vector field, where the vector field on each triangle face has a constant value orthogonal to the face normal~\cite{deGoes2015}. However, this approach is not suitable for our problem. In particular, after the mesh optimization, we need to trace along the vector field to derive polylines that approximate the rulings. A piecewise constant representation may be insufficient for capturing the variation of ruling directions within each face to allow for accurate tracing of the rulings, as shown in Fig.~\ref{fig:rep-rfileds}. Here we show a ruled surface discretized as a triangle mesh, where the original rulings are a smooth family of straight lines connecting the upper and lower boundary curves (a subset of the original rulings are shown in brown). We compute a piecewise constant vector field on the mesh to approximate the ruling directions, where the constant vector direction on each face is equal to the projected direction of the original ruling whose projection on the face passes through the face centroid. Then, we start from a point at the bottom curve and trace a polyline on the mesh (displayed in green) according to the piecewise constant vector field. We also show the original ruling that starts from the same point at the bottom (displayed in red). We can observe a notable deviation between the traced polyline and the original ruling at the top boundary, indicating an insufficient accuracy from the ruling direction approximation using the piecewise constant vector field. 

\begin{figure}[t]
    \centering
    \includegraphics[width=\linewidth]{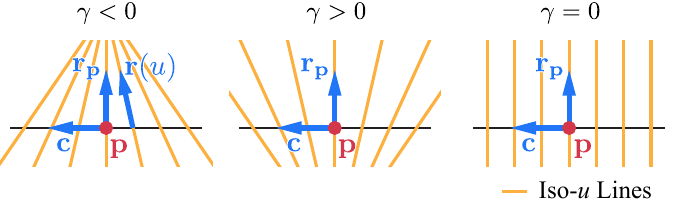}
    \caption{With our local ruled surface model in Eq.~\eqref{eq:LocalModel}, the iso-$u$ lines are straight line segments that approximate the rulings around the point $\mathbf{p}$, whereas the parameter $\gamma$ captures the local variation of the rulings.}
    \label{fig:RulingVariations}
\end{figure}

\begin{figure}[t]
    \centering
    \includegraphics[width=0.65\linewidth]{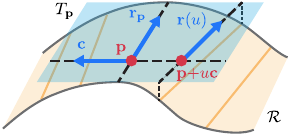}
    \caption{A schematic illustration of the local parameterization in Eq.~\eqref{eq:LocalParameterization}, which is constructed by projecting the adjacent rulings of a point $\mathbf{p}$ onto its tangent plane $\tangentplane{\mathbf{p}}$.}
    \label{fig:TangentProjection}
\end{figure}

To better approximate the ruling directions, our key idea is to construct a local $(u,v)$-parameterization on each face, such that the iso-$u$ lines are straight lines corresponding to the rulings on the face, with an additional parameter $\gamma$ controlling the local variation of the rulings (see Fig.~\ref{fig:RulingVariations}).
We derive the parameterization based on the following observation.
For a point $\mathbf{p} \in \mathbb{R}^3$ on a ruled surface $\mathcal{R}$, let $\mathbf{r}_{\mathbf{p}} \in \mathbb{R}^3$ be the unit direction vector of the ruling containing $\mathbf{p}$. Then we can project the surrounding rulings onto the tangent plane $\tangentplane{\mathbf{p}}$ at $\mathbf{p}$, and obtain a parametric form of the tangent plane as a local approximation of $\mathcal{R}$ (see Fig.~\ref{fig:TangentProjection}): 
\begin{equation}
\mathbf{q}(u, v) = \mathbf{p} +  u \mathbf{c}  + v  \mathbf{r}(u), \qquad u,v \in \mathbb{R},
\label{eq:LocalParameterization}
\end{equation}
where $\mathbf{c} \in \mathbb{R}^3$ is a unit vector in the tangent space $T_\mathbf{p}$ and is orthogonal to $\mathbf{r}_{\mathbf{p}}$; $\mathbf{r}(u) \in \mathbb{R}^3$ is the unit direction vector for the projected ruling that contains the point $\mathbf{p} + u \mathbf{c}$ (hence $\mathbf{r}(u) \in T_\mathbf{p}$), and $\mathbf{r}(0) = \mathbf{r}_{\mathbf{p}}$.
We can further simplify~\eqref{eq:LocalParameterization} by replacing $\mathbf{r}(u)$ with its first-order Taylor approximation:
\begin{equation}
    \mathbf{r}(u) \approx \mathbf{r}(0) + u  \mathbf{r}'(0) = \mathbf{r}_{\mathbf{p}} + u  \mathbf{r}'(0).
    \label{eq:TaylorApproximation}
\end{equation}
Since $\|\mathbf{r}(u)\| = 1$, the derivative $\mathbf{r}'(0)$ satisfies $\mathbf{r}'(0) \cdot \mathbf{r}(0) = 0$. Moreover, as $\mathbf{q}(u,v)$ is in the tangent plane of $\mathbf{p}$, we have $\mathbf{r}'(0) \in T_\mathbf{p}$. Thus, $\mathbf{r}'(0)$ must be linearly dependent with $\mathbf{c}$, i.e.,
\begin{equation}
    \mathbf{r}'(0) = \gamma \mathbf{c}, \qquad \gamma \in \mathbb{R}.
    \label{eq:LinearDependence}
\end{equation}
We then have a parametric local approximation of surface $\mathcal{R}$ at $\mathbf{p}$:
\begin{equation}
\overline{\mathbf{q}}(u, v) = \mathbf{p} +  u  \mathbf{c}  + v  (\mathbf{r}_{\mathbf{p}}  +  u \gamma \mathbf{c}). 
\label{eq:LocalModel}
\end{equation}
With this model, the iso-$u$ lines are straight lines that locally approximate the rulings around the point $\mathbf{p}$, while the parameter $\gamma$ controls the local ruling variation as follows (see Fig.~\ref{fig:RulingVariations}): with $\gamma < 0$, the rulings converge toward each other on the side with $v > 0$; with $\gamma > 0$, the rulings converge on the side with $v < 0$; if $\gamma = 0$, then the rulings are parallel and we have a constant ruling direction.

On a triangle mesh, we consider each face as a tangent plane associated with its centroid, and adopt the local model in Eq.~\eqref{eq:LocalModel} as a discrete representation of the rulings on the face. 
Specifically, for a face $f$, let $\facevtx{f}{1}, \facevtx{f}{2}, \facevtx{f}{3} \in \mathbb{R}^3$ be its three vertices according to a pre-defined orientation. 
We assign to the face centroid $\mathbf{o}_f$ a unit ruling direction $\facerule_f$, represented as a linear combination of two edge vectors $\faceedge{f}{1} = \facevtx{f}{2} - \facevtx{f}{1}$ and $\faceedge{f}{2} = \facevtx{f}{3} - \facevtx{f}{1}$ with scalar parameters $\rulcoefone_f, \rulcoeftwo_f \in \mathbb{R}$:
\begin{equation}
    \facerule_f = (\rulcoefone_f \faceedge{f}{1} + \rulcoeftwo_f \faceedge{f}{2})/{\|\rulcoefone_f \faceedge{f}{1} + \rulcoeftwo_f \faceedge{f}{2}\|}.
    \label{eq:FaceRuling}
\end{equation}
By construction, $\facerule_f$ is orthogonal to the face normal $\mathbf{n}_f = \frac{\faceedge{f}{1} \times \faceedge{f}{2}}{\|\faceedge{f}{1} \times \faceedge{f}{2}\|}$. 
We also compute a unit vector 
\begin{equation}
\mathbf{c}_f = \mathbf{n}_f \times \facerule_f
\label{eq:RulingOrthoDir}
\end{equation}
that belongs to the tangent space and is orthogonal to   $\facerule_f$. Then, applying Eq.~\eqref{eq:LocalModel} at $\mathbf{o}_f$, we obtain a local parameterization of $f$ that encodes the rulings on the face:
\begin{equation}
{\mathbf{q}}_f(u, v) = \mathbf{o}_f +  u~\mathbf{c}_f  + v  (\facerule_f  +  u~\gamma_f~\mathbf{c}_f), 
\label{eq:LocalFaceModel}
\end{equation}
where $\gamma_f \in \mathbb{R}$ is a ruling variation parameter.
Using Eq.~\eqref{eq:LocalFaceModel}, the ruling direction (without normalization) at any point $\facept$ on the face can be computed as (see Appendix~\ref{sec:RulingDiretionDerivation}):
\begin{equation}
    \facepointrule_f({\facept}) =  (1 + \gamma_f x_{\mathbf{s}}) \facerule_f  + \gamma_f y_{\mathbf{s}} \mathbf{c}_f,
    \quad\text{if}~1 + \gamma_f x_{\mathbf{s}}  \neq 0,
    \label{eq:PointRuilngVector}
\end{equation}
where $x_{\mathbf{s}} = (\facept - \mathbf{o}_f) \cdot \facerule_f$ and $y_{\mathbf{s}} = (\facept - \mathbf{o}_f) \cdot \mathbf{c}_f$ are local coordinates of $\facept$ with respect to the origin $\mathbf{o}_f$ and axes ($\facerule_f$,$\mathbf{c}_f$). Moreover, it can be shown that the condition $1 + \gamma_f x_{\mathbf{s}} \neq 0$ is satisfied on the whole face if the following condition is satisfied at the three vertices (see Appendix~\ref{sec:RulingConditionProof} for a proof):
\begin{equation}
    1 + \gamma_f~x_{\facevtx{f}{i}} > 0, \quad \text{for}~i =1, 2, 3.
    \label{eq:gammaCondition}
\end{equation}
Thus, we enforce \eqref{eq:gammaCondition} as a hard constraint for the parameter $\gamma_f$.

To summarize, on each mesh face we represent the ruling direction field using Eq.~\eqref{eq:PointRuilngVector}, where the parameters $(\rulcoefone_f, \rulcoeftwo_f)$ and $\gamma_f$ encode the direction at the centroid and the local variation of directions respectively,
and $\gamma_f$ is subject to the constraint~\eqref{eq:gammaCondition}.
By construction, an integral curve of the vector field~\eqref{eq:PointRuilngVector} within the face is a straight line segment. Thus, tracing the vector field across the mesh surface results in a polyline approximating a true ruling on the underlying smooth surface. With properly chosen parameters, this polyline can provide a more accurate approximation than tracing along a piecewise constant vector field, as shown in Fig.~\ref{fig:rep-rfileds}. 
Moreover, as will be shown later in Fig.~\ref{fig:test-replace}, the additional degrees of freedom offered by our representation enable more effective optimization of the piecewise ruled surface.

\subsubsection{Optimization formulation}
Our optimization is based on the following key observation: on a ruled surface, the ruling direction at each point is an asymptotic direction with zero normal curvature~\cite{flory2013ruled}; moreover, being a straight line that lies on the surface, it must also be a geodesic curve. In the following, we propose a formulation that induces these properties on the optimized surface while enforcing its closeness to the target shape.

\paragraph{Geodesic condition}
With the above observation, we require the integral curves of our ruling direction field to be as close to discrete geodesics as possible. We adopt the following conditions from~\cite{Polthier1998} for discrete geodesics on triangle meshes:
\begin{enumerate}[leftmargin=*,label=({\alph*})]
    \item Inside each face, a discrete geodesic is a straight line segment.
    \item For a discrete geodesic crossing the interior of a mesh edge, if the two adjacent faces of this edge are unfolded around the edge into a common plane, then the geodesic segments on the two faces are unfolded into a common line segment (see the inset figure below).
\end{enumerate}

\begin{wrapfigure}{r}{0.37\columnwidth}
	\vspace*{-0.8em}
	\hspace*{-1em}
	\centering
	\includegraphics[width=0.43\columnwidth]{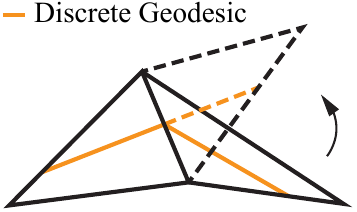}
	\vspace*{-1em}
\end{wrapfigure}
\noindent Our ruling direction field already satisfies Condition~(a) by construction. For Condition~(b), let $f_i, f_j$ be two triangle faces sharing a common edge $e_{ij}$ with end points $\mathbf{v}_1^{i,j}, \mathbf{v}_2^{i,j}$, and let $\mathbf{v}_3^{i,j}, \mathbf{v}_4^{i,j}$ be the remaining vertices from $f_i$ and $f_j$ respectively. We choose 
\begin{equation}
\mathbf{B}_{i,j}^i = [\mathbf{e}_{ij}, \mathbf{e}_{ij} \times \mathbf{n}_i], \qquad \mathbf{B}_{i,j}^j = [\mathbf{e}_{ij}, \mathbf{e}_{ij} \times \mathbf{n}_j]
\label{eq:AdjFaceBases}
\end{equation}
as the local bases for the two faces, where $\mathbf{e}_{ij} = \frac{\mathbf{v}_2^{i,j} - \mathbf{v}_1^{i,j}}{\|\mathbf{v}_2^{i,j} - \mathbf{v}_1^{i,j}\|}$ is the unit edge vector, and $\mathbf{n}_i, \mathbf{n}_j$ are the unit face normals of $f_i$ and $f_j$ respectively.\
$\mathbf{B}_{i,j}^i$ and $\mathbf{B}_{i,j}^j$ will align after the two faces are unfolded into a common plane.  
If an integral curve intersects with $e_{ij}$ at an interior point $\mathbf{q}$, then its two segments on $f_i$ and $f_j$ will have unit directions
\[
\segmentdir_i = \facepointrule_{f_i}(\mathbf{q})/\|\facepointrule_{f_i}(\mathbf{q})\|, \qquad
\segmentdir_j = \facepointrule_{f_j}(\mathbf{q})/\|\facepointrule_{f_j}(\mathbf{q})\|
\]
respectively according to Eq.~\eqref{eq:PointRuilngVector}. We use the following function to compare the local coordinates of $\segmentdir_i$ and $\segmentdir_j$ in the bases $\mathbf{B}_{i,j}^i$ and $\mathbf{B}_{i,j}^j$ respectively: 
\[
    \edgegeoderrfunc^{i,j}(\mathbf{q})
    = \left\|(\mathbf{B}_{i,j}^i)^T \segmentdir_i - (\mathbf{B}_{i,j}^j)^T \segmentdir_j \right\|^2.
\]
Note that the above measure can be seen as a discrete geodesic curvature of the integral curve at $\mathbf{q}$ and indicates its violation of Condition~(b).
To enforce Condition~(b) along the whole edge $e_{ij}$, we evenly sample $k$ interior points of $e_{ij}$ ($k=3$ in our implementation), and define an error term for $e_{ij}$ as
\begin{equation}
    \edgegeoderrfunc(e_{ij})
    = \sum\nolimits_{\mathbf{q} \in \mathcal{S}_{ij}} \edgegeoderrfunc^{i,j}(\mathbf{q}),
    \label{eq:Geod}
\end{equation}
where $\mathcal{S}_{ij}$ denotes the set of sample points.

It should be noted that for conventional piecewise constant vector fields, there exist alternative formulations to make their integral curves approximately geodesic, such as the curl-free condition from~\cite{Vekhter2019} and the divergence-free condition from~\cite{Verhoeven2022}. However, the condition from~\cite{Vekhter2019} requires the vector field directions on two adjacent faces to be symmetrical with respect to their shared edge when unfolded into a common plane, which is more restrictive than our geodesic condition (see Fig.~\ref{fig:test-replace} for an example). Meanwhile, the divergence-free condition from~\cite{Verhoeven2022} is formulated on vertices and is not directly compatible with our edge-based sparsity optimization. Therefore, we choose our current formulation to enforce the geodesic condition.

\paragraph{Curvature condition} 
As the ruling direction is also an asymptotic direction, it induces conditions on the local curvature of the underlying surface. Let $\pdir_1, \pdir_2$ be the principal directions at a point $\mathbf{p}$ on a smooth surface, and $\pcurv_1, \pcurv_2$ be the corresponding principal curvatures. Then, the normal curvature along a tangent direction $\tangentdir$ at $\mathbf{p}$ can be computed as~\cite{do2016differential}:
\[
\normalcurv{}(\tangentdir)
=
\pcurv_1 \cos^2 \theta + \pcurv_2 \sin^2 \theta, 
\]
where $\theta$ is the angle between $\pdir_1$ and $\tangentdir$.
If $\tangentdir$ is an asymptotic direction (i.e., $\normalcurv{}(\tangentdir) = 0$), then $\pcurv_1$ and $\pcurv_2$ must satisfy
\begin{equation}
    \pcurv_1 = \curvaturecoef \sin^2 \theta, \quad 
    \pcurv_2 = - \curvaturecoef \cos^2 \theta,
    \label{eq:PCurvatures}
\end{equation}
for a certain $\curvaturecoef \in \mathbb{R}$. 
We use this condition to parameterize the principal curvatures and principal directions on mesh faces.
Specifically,
\begin{wrapfigure}{r}{0.39\columnwidth}
\vspace*{-0.8em}
\hspace*{-2.3em}
\centering
    \includegraphics[width=0.45\columnwidth]{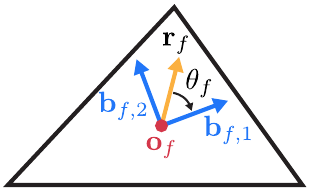}
    \vspace*{-1em}
\end{wrapfigure}
on each face $f$, we introduce a variable $\theta_f \in \mathbb{R}$ for the signed angle from the ruling direction $\facerule_f$ (defined in Eq.~\eqref{eq:FaceRuling}) to the principal direction $\pdir_{f,1}$ at the centroid $\centroid_f$ (see the inset figure). Then the principal directions $\pdir_{f,1}, \pdir_{f,2}$ at $\centroid_f$ can be represented as 
\begin{equation}
    \pdir_{f,1} = \cos \theta_f~\facerule_f + \sin \theta_f~\mathbf{c}_f, 
    \quad
    \pdir_{f,2} = -\sin \theta_f~\facerule_f + \cos \theta_f~\mathbf{c}_f,
    \label{eq:FacePrincipalDir}
\end{equation}
where $\mathbf{c}_f$ is defined in Eq.~\eqref{eq:RulingOrthoDir}. We also introduce a variable $\curvaturecoef_f \in \mathbb{R}$ and apply Eq.~\eqref{eq:PCurvatures} to represent the principal curvatures at $\centroid_f$ as
\begin{equation}
\pcurv_{f,1} = \curvaturecoef_f \sin^2 \theta_f, \quad \pcurv_{f,2} = - \curvaturecoef_f \cos^2 \theta_f.
\label{eq:FacePrincipalCurv}
\end{equation}
Using $\pdirbases_f = [\pdir_{f,1}, \pdir_{f,2}] \in \mathbb{R}^{3 \times 2}$ as the local bases for the tangent space $T_{\mathbf{o}_f}$ at $\centroid_f$, the second fundamental tensor $\sfftensor_f : T_{\mathbf{o}_f} \mapsto T_{\mathbf{o}_f}$ at $\centroid_f$ has the following form:
\[
\sfftensor_f = \text{diag}~(\pcurv_{f,1}, \pcurv_{f,2}).
\]
$\sfftensor_f$ maps each tangent vector $\mathbf{t} \in T_{\mathbf{o}_f}$ to the variation of normal along $\mathbf{t}$, and should be consistent with surface normals around $f$. Thus, for a face $f'$ that shares a common edge with $f$, we use the following term to measure the consistency between $\sfftensor_f$ and the normals at $f$ and $f'$:
\begin{equation}
    \sfffunc{f}{f'} = \|\sfftensor_f \mathbf{B}_f^T \geoddispvec{f}{f'}
    - \mathbf{B}_f^T (\mathbf{n}_{f'} - \mathbf{n}_f)\|^2\mathbin{/}\|\geoddispvec{f}{f'}\|^2.
    \label{eq:NormalConsistency}
\end{equation}

\begin{wrapfigure}{r}{0.37\columnwidth}
	\vspace*{-0.8em}
	\hspace*{-2.3em}
	\centering
	\includegraphics[width=0.43\columnwidth]{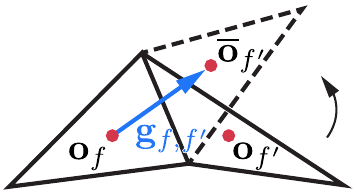}
	\vspace*{-1em}
\end{wrapfigure}
\noindent Here $\mathbf{n}_f, \mathbf{n}_{f'} \in \mathbb{R}^3$ are the unit face normals of $f$ and $f'$, respectively.
$\geoddispvec{f}{f'} = \overline{\mathbf{o}}_{f'} - \mathbf{o}_f$ is a tangent vector in $T_{\mathbf{o}_f}$ indicating the geodesic displacement between the centroids of $f$ and $f'$, where $\overline{\mathbf{o}}_{f'}$ is the centroid of $f'$ after it is unfolded into the same plane as $f$ around their common edge (see the inset figure).
The denominator in Eq.~\eqref{eq:NormalConsistency} prevents the optimization from reducing $\sfffunc{f}{f'}$ regardless of $\sfftensor_f$ by simply moving both faces to almost parallel while pulling their centroids closer.
Then, for each interior edge $e_{ij}$, we introduce a term to enforce consistency between the curvature information on its two adjacent faces $f_i$, $f_j$:
\[
    \sffterm(e_{ij}) = \sfffunc{f_i}{f_j} + \sfffunc{f_j}{f_i}. 
\]

\paragraph{Group-sparsity optimization}
For a mesh edge $e_{ij}$ inside a ruled surface patch, the terms $\edgegeoderrfunc(e_{ij})$ and $\sffterm(e_{ij})$ should both have small values. On the other hand, for an edge on the seam between two patches, at least one of the terms may have a large value due to the discontinuity of ruling directions and/or curvature information. If we simply minimize the sum of these terms over all edges, then such least-squares optimization will penalize large term values everywhere and may prevent a desirable piecewise ruled shape from emerging. 
To address this issue, we adopt a group-sparsity formulation~\cite{bach2012optimization}: for each interior mesh edge $e_{ij}$, we use the Welsch's function~\cite{holland1977robust} as a robust norm and apply it to the combined error
\begin{equation}
\combedgeerr(e_{ij}) = \edgegeoderrfunc(e_{ij})+\sffterm(e_{ij}),
\label{eq:CombinedErr}
\end{equation}
resulting in the following sparsity term for the edge errors:
\begin{equation}
    \welschterm = \sum\nolimits_{e_{ij} \in \mathcal{E}_I} \psi_{\nu}\left(\sqrt{\combedgeerr(e_{ij})}\right),
    \label{eq:WelschTerm}
\end{equation}
where $\psi_{\nu}$ is the Welsch's function defined as
\[
\psi_{\nu}(x) = 1 - \exp\left(-\frac{x^2}{2 \nu^2}\right),
\]
and $\nu > 0$ is a user-specified parameter. $\psi_{\nu}(\cdot)$ is bounded from above and insensitive to large $x$ values due to its small derivatives at such values. Meanwhile, it effectively penalizes the magnitude of $x$ when $x$ is closer to zero. Thus  $\welschterm$ can help reduce the value of $\combedgeerr$ for edges in the interior of a ruled surface patch while allowing for edges with large errors on the seams between patches, facilitating our optimization towards a piecewise ruled shape.

\paragraph{Shape closeness}
To align with the target shape, we introduce a closeness term that penalizes the deviation from each mesh vertex position $\mathbf{v}$ to the target:
\begin{equation}
    \closeterm (\mathbf{v}) = \| \mathbf{v} - P(\mathbf{v}) \|^2,
    \label{eq:Closeness}
\end{equation}
where $P(\cdot)$ denotes the closest-point projection onto the target surface (for interior vertices) or its boundary (for boundary vertices).

\paragraph{Barrier Terms}
We also introduce a barrier term to ensure the variation parameter $\gamma_f$ on each face $f$  satisfies the condition~\eqref{eq:gammaCondition}:
\begin{equation}
\barrierterm(f) =  \sum\nolimits_{i=1}^3 ({1 + \gamma_f~x_{\facevtx{f}{i}}})^{-2}.
\end{equation}
$\barrierterm(f)$  will approach infinity when $1 + \gamma_f~x_{\facevtx{f}{i}}$ approaches zero. Thus, if we start from variable values that satisfy Eq.~\eqref{eq:gammaCondition} and employ a numerical solver that monotonically decreases the target function, the term will prevent $\gamma_f$ from reaching the boundary of its feasible set and ensure the condition in Eq.~\eqref{eq:gammaCondition} remains satisfied.

\paragraph{Regularization}
We use regularizer terms to ensure smooth deformation and avoid degenerate triangles. 
For smoothness, we use a Laplacian term for the displacement of each vertex ${v}$
\begin{equation}
\displapterm(v) = \left\| (\mathbf{v} - \mathbf{v}^{(0)}) - \frac{1}{|\mathcal{N}(v)|}\sum\nolimits_{v_j \in \mathcal{N}(v)} (\mathbf{v}_j - \mathbf{v}_j^{(0)}) \right\|^2,
\end{equation}
where $\mathcal{N}(v)$ denotes the set of adjacent vertices (for an interior vertex) or the set of adjacent boundary vertices (for a boundary vertex), and $\mathbf{v}^{(0)}$ denotes the vertex position on the initial mesh.

To avoid degenerate triangles, we use the following term on each edge $e$ to prevent excessive changes in edge length:
\begin{equation}
\edgelengthterm(e) = ( {\|\mathbf{h}_e\| - \|\mathbf{h}_e^{(0)}\|})^2 \mathbin{/} \|\mathbf{h}_e^{(0)}\|^2,
\end{equation}
where $\mathbf{h}_e$ and $\mathbf{h}_e^{(0)}$ are the edge vector of $e$ in the current mesh and the initial mesh, respectively.

\paragraph{Overall Formulation} Combining the terms presented above, we obtain the target function for our optimization:
\begin{equation}
\begin{aligned}
    E =~ & \welschterm + \closeweight \sum\nolimits_{v \in \mathcal{V}} \closeterm(v) + \barrierweight \sum\nolimits_{f \in \mathcal{F}} \barrierterm(f)\\
    & + \lapweight \sum\nolimits_{v \in \mathcal{V}} \displapterm(v)  + \edgelengthweight \sum\nolimits_{e \in \mathcal{E}} \edgelengthterm(e),
\end{aligned}
\label{eq:JointOptTargetFunction}
\end{equation}
where $\mathcal{V}, \mathcal{E}, \mathcal{F}$ denote the set of mesh vertices, mesh edges, and mesh faces respectively, and $\closeweight, \lapweight, \edgelengthweight, \barrierweight$ are user-specified weights. 
The variables include all the mesh vertex positions, as well as the parameters $\{(\rulcoefone_f, \rulcoeftwo_f,\gamma_f)\}$ and $\{(\curvaturecoef_f, \theta_f)\}$ that represent the ruling direction field and local curvature information on each face.

We solve this problem numerically with an L-BFGS solver, using automatic differentiation to evaluate the gradient. 
We adopt a line search strategy that rejects any stepsize violating the hard constraints encoded by the barrier functions, ensuring all constraints are always satisfied.
To improve efficiency, the values and gradients of different target function terms are computed in parallel. 

It is worth noting that the parameter $\nu$ in the Welsch's function controls its robustness: with a $\nu$ closer to zero, the function is less sensitive to large error values, which would benefit the emergence of seams; however, if we use a very small value of $\nu$ during the whole process, $\welschterm$ would have small gradients on most surface areas and may fail to steer the mesh toward a desirable shape. Thus, following existing methods that use the Welsch's function as a robust norm~\cite{Zhang2022Fast}, we start with a large value $\nu_{\max}$ for $\nu$ and gradually decrease $\nu$ during optimization by repeatedly reducing it by half, until $\nu$ reaches a minimum value $\nu_{\min}$.

\subsubsection{Vector field initialization}
\label{sec:InitVectorField}
\begin{figure}[t!]
    \centering
    \includegraphics[width=\linewidth]{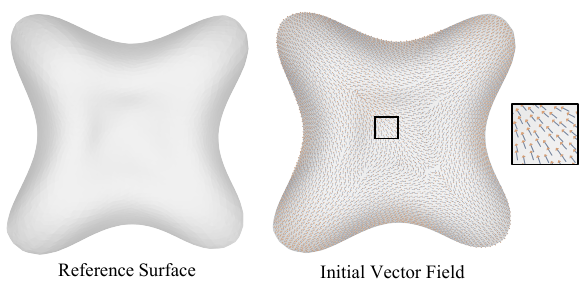}
    \caption{An initial vector field computed using the method in Sec.~\ref{sec:InitVectorField}.}
    \label{fig:InitVectorField}
\end{figure}

Since our optimization problem is nonconvex, a proper initialization of the ruling direction field is necessary for the solver to produce desirable results. 
To this end, we initialize all ruling variation variable $\{\gamma_f\}$ to zero, and construct the initial ruling directions $\{\facerule_f\}$ at the face centroids as a smooth vector field (see Fig.~\ref{fig:InitVectorField} for an example).

In~\cite{flory2013ruled}, it is suggested that the initial ruling directions should align with the asymptotic directions across the surface. However, this is ineffective for surface regions of positive Gaussian curvature, as there is no asymptotic direction there. Moreover, at a point of negative Gaussian curvature, there are more than one asymptotic directions, and we need to choose one of them in a globally consistent way to ensure the overall smoothness of the resulting direction field.
To address these challenges, we first generalize the condition in~\cite{flory2013ruled}: On each mesh face, we require the initial ruling direction to align with a tangent direction with the smallest magnitude of normal curvature. This is based on the following observation: At a surface point $\mathbf{p}$, the normal curvature along a tangent direction $\mathbf{t}$ is the extrinsic curvature of the geodesic curve that passes through $\mathbf{p}$ along $\mathbf{t}$; therefore, among all geodesic curves passing through $\mathbf{p}$, the one along the tangent direction with the smallest magnitude of normal curvature has the smallest local deviation from a straight line passing through $\mathbf{p}$ along the same tangent direction, making it suitable for our initialization.
It is worth noting that at a point with non-positive Gaussian curvature, our condition is equivalent to the one proposed in~\cite{flory2013ruled}, since the normal curvature attains the minimum magnitude of zero along the asymptotic directions. Therefore, our condition is a generalization of the one in~\cite{flory2013ruled}. 
Meanwhile, at a point with positive Gaussian curvature, our condition requires the initial direction to align with a principal direction with the smallest magnitude of principal curvature. To define the target directions, we use the method of~\cite{rusinkiewicz2004estimating} to determine the principal directions and principal curvatures on each face.

\begin{figure}[t!]
    \centering
    \includegraphics[width=0.95\linewidth]{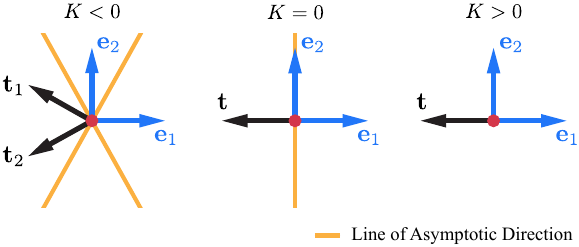}
    \caption{The definition of vector $\mathbf{t}$ used in Eq.~\eqref{eq:InitTargetAlignTerm} depending on the sign of the Gaussian curvature $K$. Here, $\mathbf{e}_1$ and $\mathbf{e}_2$ are principal directions, with their corresponding principal curvatures $\kappa_1$ and $\kappa_2$ satisfying $|\kappa_1| \geq |\kappa_2|$. With a negative Gaussian curvature, there are two vectors $\mathbf{t}_1$ and $\mathbf{t}_2$ orthogonal to the two lines of asymptotic direction, respectively. Otherwise, there is a single $\mathbf{t}$ parallel with $\mathbf{e}_1$.}
    \label{fig:TargetDirections}
\end{figure}

To compute a smooth vector field $\{\facerule_f\}$ that meets the alignment conditions, we first construct a local orthonormal basis $\faceorthobases_f \in \mathbb{R}^{3 \times 2}$ for each face $f$ and represent  $\facerule_f$ as $\facerule_f = \faceorthobases_f \initlocalcoord_f$ with local coordinates $\initlocalcoord_f \in \mathbb{R}^2$. Then we solve an optimization problem for $\{\initlocalcoord_f\}$:
\begin{equation}
    \min_{\{\initlocalcoord_f\}}
    ~\smoothterm + \unitweight \unitterm + ~\npfacealignweight \sum_{f \in \mathcal{F}_{\textrm{np}}}  \initalignterm (\initlocalcoord_f)
    ~+ \pfacealignweight \sum_{f' \in \mathcal{F}_{\textrm{p}}} \initalignterm(\initlocalcoord_{f'}),
    \label{eq:InitOpt}
\end{equation}
where $\unitweight$, $\npfacealignweight$ and $\pfacealignweight$ are user-specified weights.
Here, the term 
\begin{equation}
\initalignterm(\initlocalcoord_f) =  \min\nolimits_{\mathbf{t} \in \asympdirset_f} \| \initlocalcoord_f \cdot \mathbf{t}\|^2
\label{eq:InitTargetAlignTerm}
\end{equation}
measures the alignment between $\initlocalcoord_f$ and its target directions defined above, where $\asympdirset_f$ is a set of unit tangent vectors orthogonal to target direction and represented with local coordinates with respect to $\faceorthobases_f$, and $\mathcal{F}_{\textrm{np}}$, $\mathcal{F}_{\textrm{p}}$ denote the set of faces with non-positive Gaussian curvature and positive Gaussian curvature, respectively. On a face with non-negative Gaussian curvature, $\asympdirset_f$ contains only one vector along the principal direction with the largest magnitude of principal curvature. On a face with negative Gaussian curvature,  $\asympdirset_f$ contains two vectors orthogonal to the two lines of asymptotic direction, respectively (see Fig.~\ref{fig:TargetDirections}). 
Additionally,  the term $\smoothterm$ measures the smoothness of the vector field by comparing its values on each pair of adjacent faces $f_i, f_j$ after unfolding them into a common plane around their shared edge:
\[
\smoothterm =  \sum\nolimits_{e_{ij} \in \mathcal{E}_I} \| (\mathbf{B}_{i,j}^i)^T \faceorthobases_{f_i} \initlocalcoord_{f_i} - (\mathbf{B}_{i,j}^j)^T \faceorthobases_{f_j} \initlocalcoord_{f_j}\|^2,
\]
where $\mathbf{B}_{i,j}^i, \mathbf{B}_{i,j}^j$ are local bases defined in Eq.~\eqref{eq:AdjFaceBases}.
Lastly, the term 
\[
\unitterm = \sum\nolimits_{f \in \mathcal{F}} \left\|\initlocalcoord_f - \frac{\initlocalcoord_f}{\|\initlocalcoord_f\|}\right\|^2
\]
enforces a unit-length condition for the vectors $\{\initlocalcoord_f\}$ as well as $\{\facerule_f\}$. 
We solve the problem~\eqref{eq:InitOpt} using a majorization-minimization (MM) solver~\cite{lange2016mm}. Full details can be found in Appendix~\ref{sec:MMSolver}.

\subsection{Initial Piecewise Ruled Surface}
\label{sec:InitSurf}
\begin{figure}[t]
    \centering
    \includegraphics[width=\linewidth]{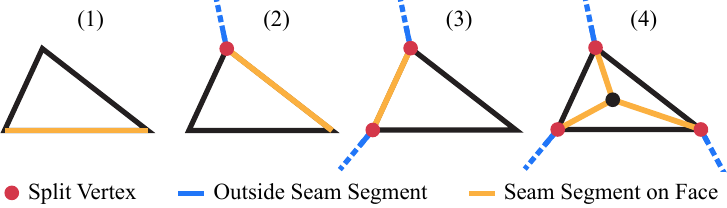}
    \caption{Construction of seam segments on a connected surface region $\facecomponent$ with a single face, according to the number of its split vertices.}
    \label{figs:sgf-subd}
\end{figure}

After the joint optimization of the mesh shape and the ruling direction field, we use the optimization result to extract polyline seams between ruled surface patches and construct initial rulings. As candidates of seam segments, we first extract a set $\candidateedgeset$ of edges whose combined error values $\combedgeerr$ defined in Eq.~\eqref{eq:CombinedErr} are greater than a threshold $\welshthreshold$. The union of these candidate edges may have complicated topologies such as small loops (e.g., see Fig.~\ref{fig:region-contract}), which are not suitable for our seam representation. Therefore, we clean up the candidates and construct simple seams as follows. 
First, we identify a set $\boundaryfaceset$ of faces 
representing the regions that require seam simplification. 
$\boundaryfaceset$ includes all faces that contain at least two edges from $\candidateedgeset$: a face with three such edges forms a small loop of candidate edges that need to be simplified, while a face with two such edges can arise from zigzag seams that would benefit from simplification. Additionally, if some edges in $\candidateedgeset$ form a loop that encloses an area smaller than a threshold $\smallareathreshold$, we add all the enclosed faces to $\boundaryfaceset$ to avoid creating a very small patch.
Candidate edges in $\candidateedgeset$ that do not belong to the faces in $\boundaryfaceset$ are directly used as part of the seams, while additional seam segments are constructed on the faces in $\boundaryfaceset$ using the following steps.
First, we construct a dual graph for $\boundaryfaceset$, where the vertices in the dual graph correspond to the faces in $\boundaryfaceset$, and two vertices are connected if their corresponding faces share a common mesh edge.
Each connected component of this dual graph corresponds to a connected surface region $\facecomponent$ consisting of faces from $\boundaryfaceset$.
We construct seam segments inside $\facecomponent$ according to candidate edges that are outside $\facecomponent$ but attached to the boundary vertices of $\facecomponent$. We call such a vertex a \emph{split vertex}. Our goal is to construct seam segments that have simple shapes and connect these split vertices, in order to form continuous seams with simple topology (see Fig.~\ref{fig:region-contract} for an example). To this end, we consider two cases:
\begin{figure}[t]
    \centering
    \includegraphics[width=0.95\linewidth]{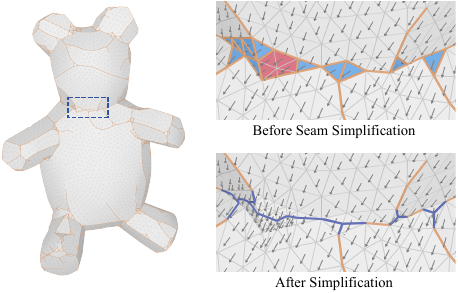}
    \caption{An example of seam simplification. Top right: we first extract candidate edges (shown in brown) for the seam segments according to the optimized ruling direction field (shown in gray), as well as areas where we need to perform seam simplification; the simplification areas include faces with at least two candidate edges (shown in blue) and small areas enclosed by candidates edges (shown in red). Bottom right: within each simplification area, we use graph cut to create simple seam segments (shown in purple) connecting the candidate edges outside the area, and further optimize the ruling direction field inside the area according to target function~\eqref{eq:PostProcessRulingFieldOpt}.}
    \label{fig:region-contract}
\end{figure}
\begin{itemize}[leftmargin=*]
    \item If  $\facecomponent$ contains a single face, then we construct a set of seam segments within the face according to the number of its split vertices. We consider four sub-cases (see Fig.~\ref{figs:sgf-subd}):
    \begin{enumerate}[leftmargin=*,label={(\arabic*)}]
        \item  If there is no split vertex, we choose the longest edge of the face as the only seam segment.
        \item If there is one split vertex, we choose the longest edge attached to this vertex.
        \item If there are two split vertices, we choose the edge between them.
        \item If there are three split vertices, we add a new vertex at the face centroid and connect it with the three face vertices to subdivide the face into three smaller faces, and then choose the edges between the three faces as the seam segments.
    \end{enumerate}
    \item If $\facecomponent$ contains more than one face, then we consider different cases according to the number of split vertices on the boundary of  $\facecomponent$:
    \begin{enumerate}[leftmargin=*,label={(\arabic*)}]
    \item If there are $\numcompbdrsegs \geq 2$ split vertices, then they divide the boundary of $\facecomponent$ into $\numcompbdrsegs$ segments. We perform a graph cut optimization to divide $\facecomponent$ into $\numcompbdrsegs$ regions, with each region attached to one boundary segment of $\facecomponent$ (see Appendix~\ref{sec:GraphCut} for details). In this way, the boundaries between the regions connect the split vertices, while the graph cut helps simplify their shapes. We then add the inter-region boundaries as seam segments.
    \item If there is only one split vertex $\singlesplitvtx$, then we find a vertex $\virtualsplitvtx$ on the boundary of $\facecomponent$ with the largest geodesic distance to $\singlesplitvtx$, and add $\virtualsplitvtx$ as a virtual split vertex. Afterward, we have two split vertices, and we use the same procedure as Case~(1) above to construct the seam segments. In this way, the resulting segments align with the overall shape of $\facecomponent$.
    \item If there is no split vertex, then we find a pair of vertices $\virtualsplitvtx_1$ and $\virtualsplitvtx_2$ on the boundary of $\facecomponent$ with the largest value of geodesic distance among all such pairs. We then add $\virtualsplitvtx_1$, $\virtualsplitvtx_2$ as virtual split vertices, and apply the procedure in Case~(1).
    \end{enumerate}
\end{itemize}

The polyline seam segments produced by the above procedure, together with the remaining candidate edges from  $\candidateedgeset$ that are outside $\boundaryfaceset$, form the full set of seams.
It is worth noting that our approach may produce not only loops of seams that fully enclose patches, but also isolated non-loop seams. We allow such non-loop seams as they enable better approximation of the target shape, such as in areas with positive Gaussian curvature surrounded by negative Gaussian curvature regions.

After the clean-up, we further optimize the ruling direction field on $\facecomponent$ by minimizing a target energy 
\begin{equation}
\sum\nolimits_{e_{ij} \in \facecompedgeset} \edgegeoderrfunc(e_{ij}),
\label{eq:PostProcessRulingFieldOpt}
\end{equation}
where $\facecompedgeset$ is the set of interior mesh edges in $\facecomponent$ that are not on the seams. As the patch topology has been determined, this optimization specifically reduces the geodesic curvature of integral curves inside the patches, which facilitates our construction of initial rulings.

\begin{figure}[t!]
    \centering
    \includegraphics[width=\linewidth]{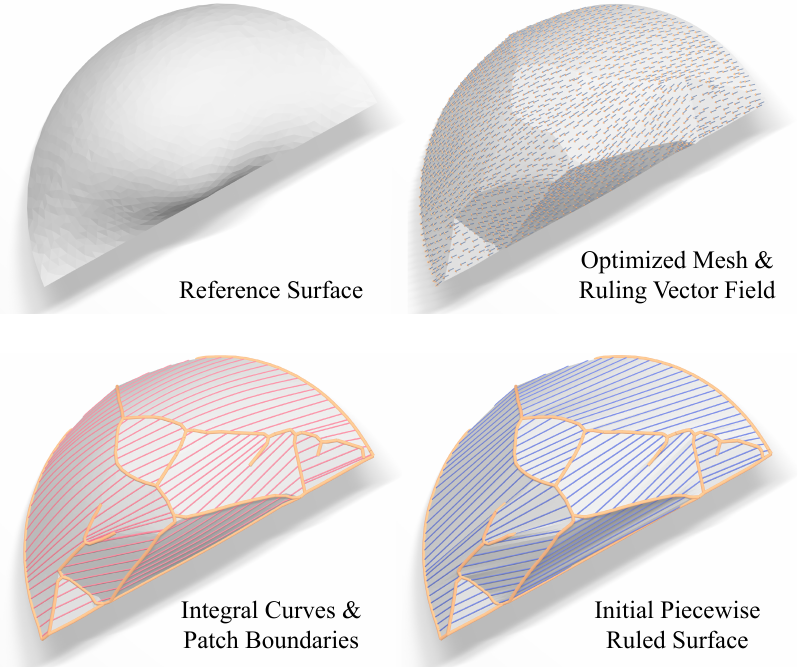}
    \caption{Construction of the initial piecewise ruled surface. Using the optimized mesh and ruling vector field (top right), we extract the initial seams and trace along the vector field to derive integral curves (bottom left). Then we connect each integral curve's two endpoints on the seams or the mesh boundary, to derive the initial rulings (bottom right).}
    \label{fig:initial-rec-mesh}
\end{figure}

Finally, we start from sample points on the optimized mesh and trace along the ruling vector field, to produce a dense set of polyline integral curves whose endpoints lie on the seams or the mesh boundary. We then connect the two endpoints of each integral curve to derive the initial rulings (see Fig.~\ref{fig:initial-rec-mesh} for an example).

\subsection{Piecewise Ruled Surface Optimization}
The initial rulings constructed in Section~\ref{sec:InitSurf} already align roughly with the target shape. As the final step, we further optimize them to improve the alignment.
Recall that the ruling endpoints lie on either the seams or the mesh boundary, which form the boundaries of the patches.   
Thus, we first represent the seam and mesh boundary polylines using the ruling endpoints as their vertices; in this way, the ruling endpoints control both the rulings and the boundary shape of the patches, facilitating our optimization formulation. We then optimize the vertex positions $\{\mathbf{b}_i\}$ of all the patch boundaries, to align the piecewise ruled surface with the target shape while ensuring the smoothness of both the patches and their boundaries.

We enforce surface alignment using the following target term that penalizes the distance to the target surface for both the patch boundaries and the rulings:
\[
    \finalalignterm
    =  \frac{\finalalignintweight}{M_1}\sum\nolimits_{\mathbf{s} \in \rulingsampleset} \alignptweight_{\mathbf{s}} \closeterm(\mathbf{s})
    +  \frac{\finalalignbdrweight}{M_2}\sum\nolimits_{\mathbf{b} \in \bdrsampleset} \alignptweight_{\mathbf{b}} \closeterm(\mathbf{b}).
\]
Here $\closeterm$ is defined in Eq.~\eqref{eq:Closeness}, and the sets $\rulingsampleset$ and $\bdrsampleset$ contain the patch boundary polyline vertices and uniform sample points on the rulings, respectively. Each sample point on a ruling is represented as a convex combination of its two endpoints with fixed weights.  $\alignptweight_{\mathbf{p}}$ is an constant weight that equals the squared mean distance from  $\mathbf{p}$ to its $k$-nearest neighbors in the set $\rulingsampleset \cup \bdrsampleset$ ($k=4$ in our implementation), computed using their initial positions.  $M_1 = \sum_{\mathbf{s} \in \rulingsampleset}\alignptweight_{\mathbf{s}}$ and $M_2 = \sum_{\mathbf{b} \in \bdrsampleset}\alignptweight_{\mathbf{b}}$ are normalization factors.
And $\finalalignintweight$, $\finalalignbdrweight$ are user-specified weights.

\begin{figure}[t]
	\centering
	\includegraphics[width=0.6\linewidth]{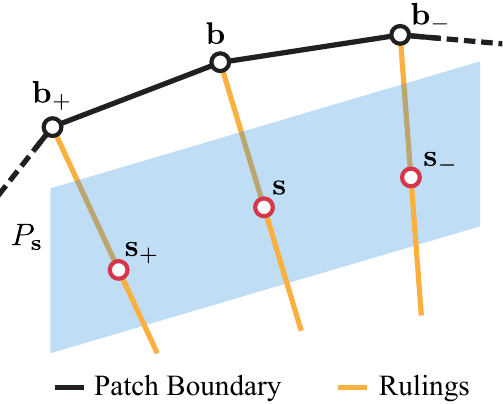}
	\caption{Definition of the points $\mathbf{b}$, $\mathbf{b}_{+}$, $\mathbf{b}_{-}$ and $\mathbf{s}$, $\mathbf{s}_{+}$, $\mathbf{s}_{-}$  for the curvature functions in Eq.~\eqref{eq:PatchNormalCurvature} and Eq.~\eqref{eq:BoundaryCurvature}, respectively.}
	\label{fig:SmoothnessTerms}
\end{figure}

Additionally, we enforce the smoothness of patch shapes by penalizing their curvature. Specifically, for each sample point $\rulingsamplept$ on a ruling, we construct a plane ${P}_{\rulingsamplept}$ 
that contains $\rulingsamplept$ and is orthogonal to the ruling, and intersect ${P}_{\rulingsamplept}$ with the two adjacent rulings to obtain points $\rulingsamplept_{+}$ and $\rulingsamplept_{-}$ (see Fig.~\ref{fig:SmoothnessTerms}). 
Then we use the function
\begin{equation}
\normalcurvfunc(\rulingsamplept) = \left\|\frac{({\rulingsamplept_{+} - \rulingsamplept}){/}{\|\rulingsamplept_{+} - \rulingsamplept\|} - (\rulingsamplept - \rulingsamplept_{-}){/}{\|\rulingsamplept - \rulingsamplept_{-}\|}}{({\|\rulingsamplept_{+} - \rulingsamplept\| + \|\rulingsamplept - \rulingsamplept_{-}\|}){/}{2}}\right\|
\label{eq:PatchNormalCurvature}
\end{equation}
to measure the magnitude of the ruled surface's discrete normal curvature at $\rulingsamplept$ along the direction orthogonal to the ruling. Based on this function, we introduce a surface smoothness term:
\[
    \rulingsmoothterm = \frac{1}{M_3} \sum\nolimits_{\rulingsamplept \in \rulingsampleset} \patchsmoothptweight_{\rulingsamplept} [\normalcurvfunc(\rulingsamplept)]^2,
\]
where $\patchsmoothptweight_{\rulingsamplept}$ is an area weight equal to the initial squared mean distance from ${\rulingsamplept}$ to its two nearest sample points on the same ruling and the closest sample point on each adjacent ruling, and $M_3 = \sum_{\rulingsamplept \in \rulingsampleset} \patchsmoothptweight_{\rulingsamplept}$.

Moreover, to enforce the smoothness of each patch boundary polyline, we evaluate the magnitude of its discrete curvature at each interior vertex $\bdrsamplept$ via 
\begin{equation}
\bdrcurvfunc(\bdrsamplept) = \left\|\frac{({\bdrsamplept_{+} - \bdrsamplept}){/}{\|\bdrsamplept_{+} - \bdrsamplept\|} - ({\bdrsamplept - \bdrsamplept_{-}}){/}{\|\bdrsamplept - \bdrsamplept_{-}\|}}{(\|\bdrsamplept_{+} - \bdrsamplept\| + \|\bdrsamplept - \bdrsamplept_{-}\|){/}2}\right\|,
\label{eq:BoundaryCurvature}
\end{equation}
where $\bdrsamplept_{+}$ and $\bdrsamplept_{-}$ are the two adjacent vertices for $\bdrsamplept$ on the polyline (see Fig.~\ref{fig:SmoothnessTerms}). Then we introduce a boundary smoothness term:
\[
    \boundarysmoothterm = \frac{1}{M_4} \sum\nolimits_{\bdrsamplept \in \bdrsampleset} \bdrsmoothptweight_{\bdrsamplept} [\bdrcurvfunc(\bdrsamplept)]^2,
\]
where $\bdrsmoothptweight_{\bdrsamplept}$ is a constant weight equal to the average initial distance from $\bdrsamplept$ to $\bdrsamplept^{-}$ and $\bdrsamplept^{+}$, and $M_4 = \sum\nolimits_{\bdrsamplept \in \bdrsampleset} \bdrsmoothptweight_{\bdrsamplept}$.

Our overall formulation is written as
\begin{equation}
    \min~ \finalalignterm + \rulingsmoothtermweight \rulingsmoothterm + \bdrsmoothtermweight \boundarysmoothterm,
    \label{eq:FinalOptProblem}
\end{equation}
where $\rulingsmoothtermweight$ and $\bdrsmoothtermweight$ are user-specified weights. We use a variant of Powell's Dogleg method~\cite{byrd1988approximate} to solve this non-linear least-squares problem.

After the optimization, we can add more rulings between existing ones if needed for an application. Since the rulings define a discrete correspondence between the boundary polylines via their endpoints, a simple approach is to interpolate the correspondence using arc-length parameterization of the boundary polyline segments between the endpoints of two adjacent rulings, and then connect the interpolated corresponding points to form new rulings. Alternatively, if higher-order smoothness is required, we could convert the boundary polylines into smooth curves using cubic spline interpolation of the polyline vertices, and connect point pairs with corresponding parameters to create new rulings. In our experiments, we do not find a noticeable difference between the two approaches since the rulings are already dense before refinement. 

%% file: experiment.tex
\section{Results}

We implement our method in C++, using the \textsc{Eigen} library~\cite{eigenweb} for linear algebra operations and OpenMP for parallelization.
We use the L-BFGS solver from \textsc{LBFGSpp}~\footnote{\url{https://github.com/yixuan/LBFGSpp/}} for the problem~\eqref{eq:JointOptTargetFunction}, and the Dogleg method from \textsc{Ceres}~\cite{Agarwal_Ceres_Solver_2022} for the problem~\eqref{eq:FinalOptProblem}.
The parameter setting of our method is provided in Appendix~\ref{eq:ParamSetting}.
All experiments are run on a PC with a 10-core Intel Core i9-10900 CPU at 2.8GHz and 32GB of RAM.

\paragraph{Comparison with other methods} 
We evaluate our method by testing it on a diverse set of freeform surfaces represented as triangle meshes, including both closed surfaces of non-zero genus and open surfaces with boundary.
Each reference surface is normalized to have unit length for its bounding box diagonal.
As no existing piecewise ruled surface approximation method works directly on mesh surfaces as far as we are aware, we instead compare with three recent methods from~\cite{Stein2018}, \cite{Zhao2022} and \cite{Zhao2023} that produce piecewise developable approximation. For each method, we use the open-source implementation\footnote{\url{https://github.com/odedstein/DevelopabilityOfTriangleMeshes}}$^{,}$\footnote{\url{hhttps://github.com/QingFang1208/DevelopApp}}$^{,}$\footnote{\url{https://github.com/mmoolee/EvoDevelop}} released by the authors.
As our goal is to approximate the given surface with a simple patch layout, we compare the results using their approximation accuracy as well as patch complexity. For approximation accuracy, we compute the average distance $\avgerr$ and maximum distance $\maxerr$ from sample points on the result surface to the reference surface.
For patch complexity, we evaluate the total length $\seamlength$ of all the seams, with a shorter length indicating a simpler patch layout. However, unlike our approach that directly constructs seams and rulings, none of the piecewise developable results contains explicit ruling information, and some of them contain no seam information either. Specifically, the results from~\cite{Stein2018} are only represented as triangle meshes without a seam representation. Additionally, although~\cite{Zhao2022} and \cite{Zhao2023} model the seam curves explicitly, we observe that their results can still contain points of large Gaussian curvature in the interior of a patch, indicating that the patch is not fully developable. To properly measure the seam length, we note the observation from~\cite{Stein2018} that on a piecewise developable surface, the Gaussian curvature must be concentrated at the seams since such high-curvature regions cannot be located in the interior of a developable patch by definition. Therefore, if both vertices of a mesh edge are in the interior of the mesh and have Gaussian curvature magnitudes larger than a threshold $\overline{\kappa}$, then we include the edge as part of the seams. Additionally, for \cite{Zhao2022} and \cite{Zhao2023}, we include all interior edges labeled by their methods as patch boundaries into the seams. To calculate Gaussian curvature for a mesh vertex, we evaluate its Gaussian curvature operator as defined in~\cite[Eq.~(9)]{meyer2003discrete}, which measures the average Gaussian curvature within the associated area of the vertex. We set the Gaussian curvature threshold to $\overline{\kappa} = 0.5$, which is a significant value since the reference surface's bounding box diagonal  has unit length.

\begin{figure*}[p]
	\centering
	\includegraphics[width=0.9\linewidth]{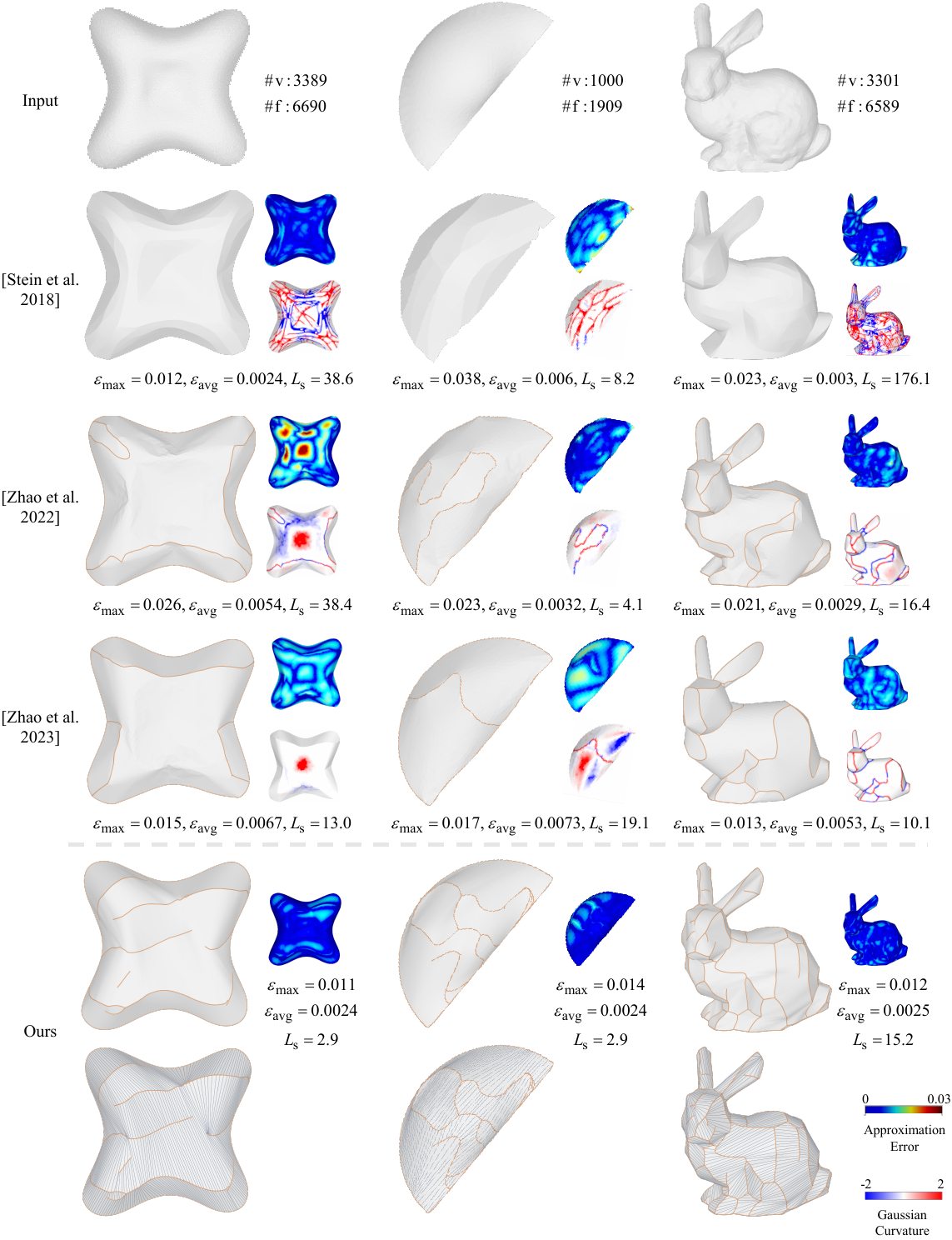}
	\caption{Comparison with \cite{Stein2018}, \cite{Zhao2022} and \cite{Zhao2023}. The color coding shows the distance from the reference surface, as well as the Gaussian curvature on the piecewise developable results.}
	\label{fig:compare}
\end{figure*}

\begin{figure*}[t]
	\centering
	\includegraphics[width=0.9\linewidth]{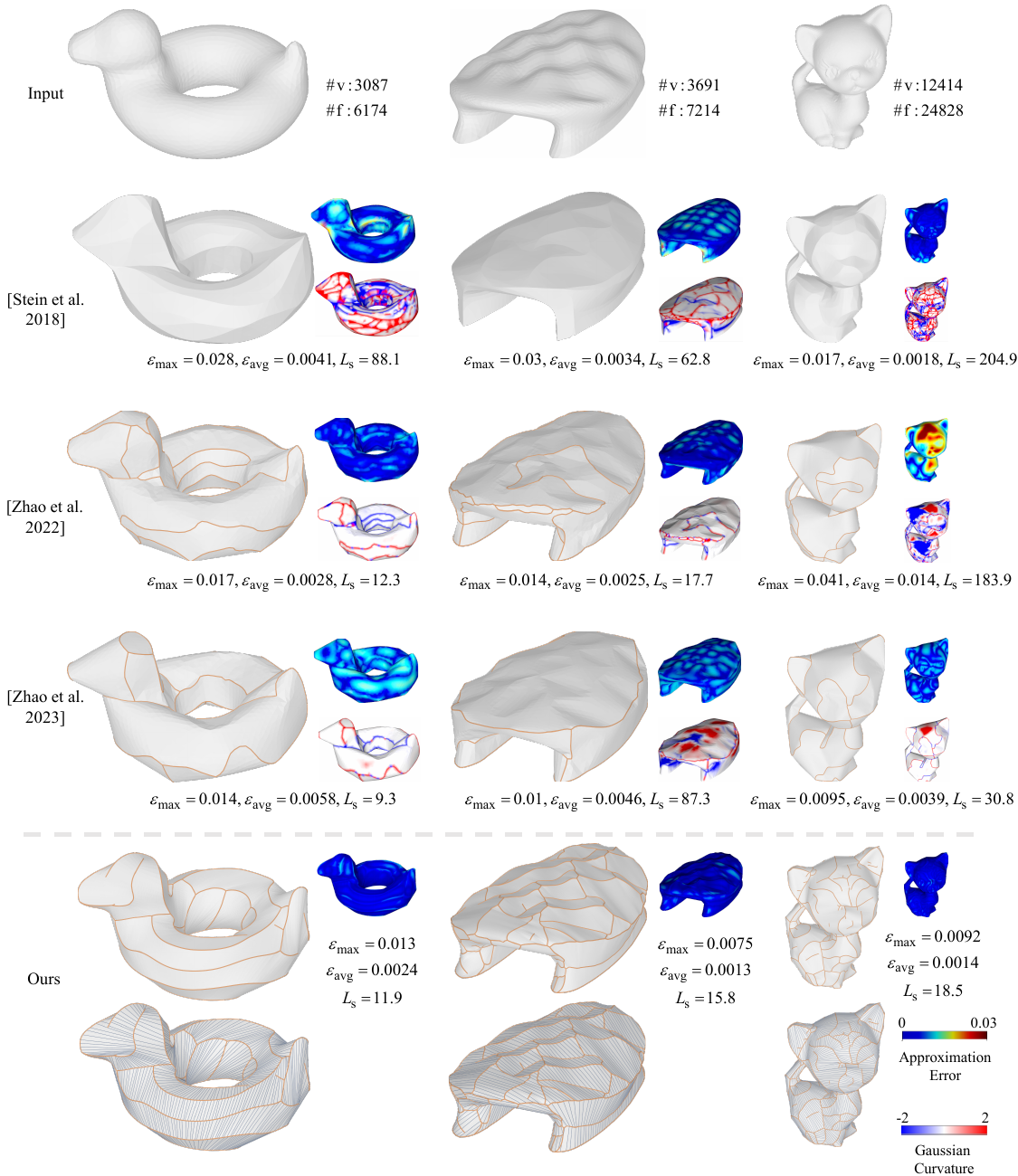}
	\caption{Comparison with \cite{Stein2018}, \cite{Zhao2022} and \cite{Zhao2023}. The color coding shows the distance from the reference surface, as well as the Gaussian curvature on the piecewise developable results.}
	\label{fig:compare1}
\end{figure*}

\begin{figure*}[t]
	\centering
	\includegraphics[width=\linewidth]{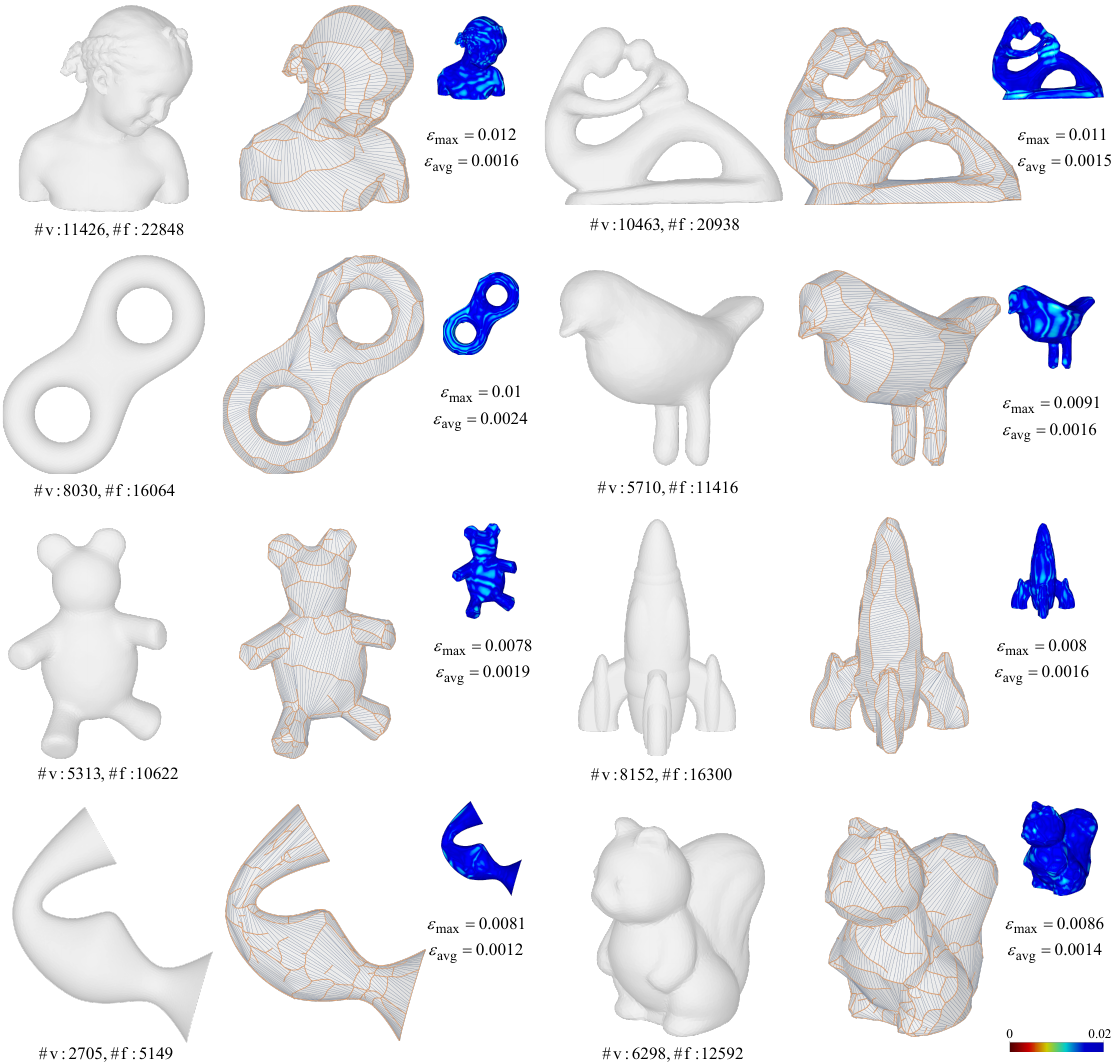}
	\caption{More results using our methods on various reference surfaces.}
	\label{fig:gallery}
\end{figure*}

\begin{figure*}[t]
	\centering
	\includegraphics[width=\linewidth]{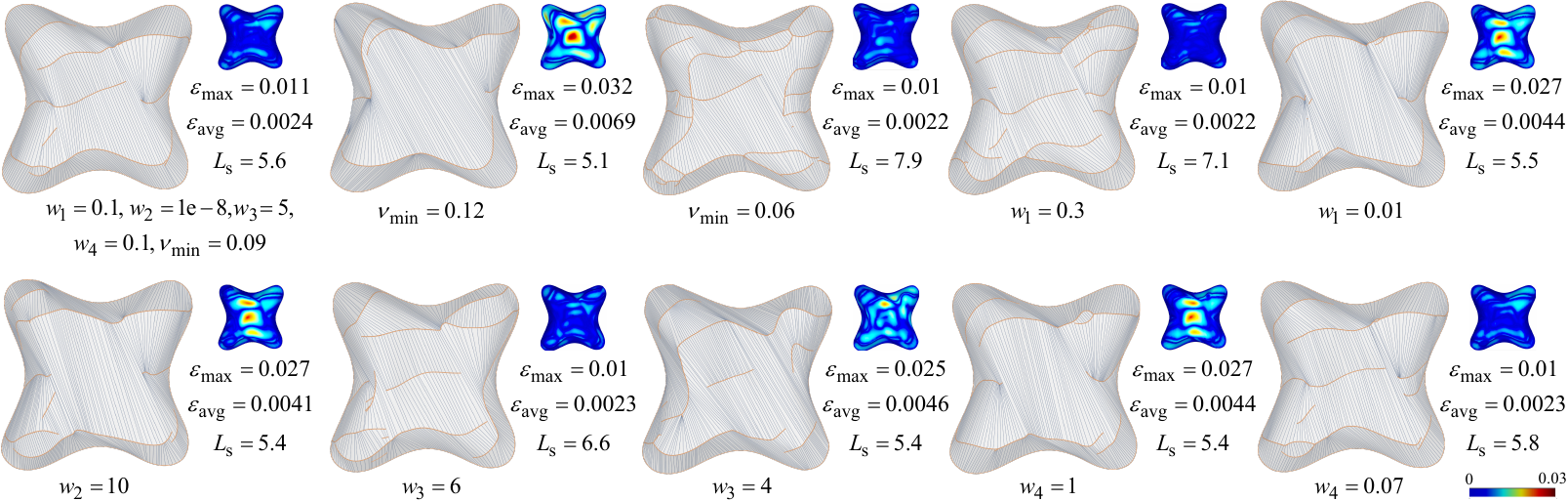}
	\caption{Influence of different parameters for the target function in Eq.~\eqref{eq:JointOptTargetFunction} on the final result. The topleft is the baseline parameter setting for the result in Fig.~\ref{fig:pipeline}. In each of the remaining subfigures, we change one parameter compared to the baseline setting, with the modified parameter value shown below the result image.}
	\label{fig:nu-comp}
\end{figure*}

Fig.~\ref{fig:compare} and Fig.~\ref{fig:compare1} compare different methods on the same set of reference surfaces. 
For each result, we display the metric values $\avgerr$, $\maxerr$ and $\seamlength$. We also use color coding to visualize the deviation from a result surface to the reference surface, as well as the Gaussian curvature of the piecewise developable results. 
For \cite{Zhao2022} and \cite{Zhao2023}, we also display the patch boundaries produced by their methods in brown.
From the metric data and the visualization, we can see that our method achieves the lowest average error and the lowest maximum error, demonstrating good approximation accuracy. 
Additionally, on models with large areas of negative Gaussian curvature (e.g., the top part of the `airport' model in the middle column of Fig.~\ref{fig:compare1}), our method has an even stronger advantage in approximation accuracy. This is not a surprise since developable surfaces can only have zero Gaussian curvature while ruled surfaces allow for non-positive Gaussian curvature. Therefore, our piecewise ruled approximation provides more flexibility for accurate approximation in negative Gaussian curvature regions compared to piecewise developable surfaces.
Moreover, our method achieves the lowest total seam length on almost all examples, except for the Bunny model in Fig.~\ref{fig:compare} (rightmost column) and the Bob modeling Fig.~\ref{fig:compare1} (leftmost column), where the method of \cite{Zhao2023} produces slightly shorter seam length but with notably higher approximation errors. 
We also note from the Gaussian curvature visualization that although both \cite{Zhao2022} and \cite{Zhao2023} produce simple layouts of patch boundaries, their patches can still contain notable regions of high Gaussian curvature in the interior where the surface is not fully developable.
Fig.~\ref{fig:gallery} shows more results of our method on reference surfaces with different shapes and topologies. In all examples, our maximum and average errors are no more than 1.5\% and 0.25\% of the reference shape's bounding box diagonal length, respectively.
Overall, the results verify the effectiveness of our method in approximating arbitrary freeform shapes using piecewise ruled surfaces with low approximation error and limited presence of seams.

\paragraph{Influence of parameters}
In Fig.~\ref{fig:nu-comp}, we demonstrate the influence of different parameters for the target function in Eq.~\eqref{eq:JointOptTargetFunction} on the final result.
We use our parameter settings for the example in Fig.~\ref{fig:pipeline} as a baseline, and vary the value of a single parameter to see how it affects the result. 
We can observe the following from Fig.~\ref{fig:nu-comp}:
\begin{itemize}[leftmargin=*]
    \item The final value $\nu_{\min}$ of the parameter $\nu$ for the Welsch's function in Eq.~\eqref{eq:WelschTerm} affects the piecewise smoothness of the optimized ruling direction field. Decreasing $\nu_{\min}$ will enforce a stronger smoothness requirement for the ruling vector field in the interior of a patch, which induces more edges with a large error value of $\combedgeerr$ that indicates seams. As a result, the final piecewise ruled surface tends to have more patches while being closer to the target surface. On the other hand, increasing the value of $\nu_{\min}$ tends to produce a simpler patch layout with a larger deviation from the target surface. Thus, $\nu_{\min}$ can be used to control the complexity of the patch layout.
    \item Increasing $\closeweight$ can reduce the approximation error at the cost of introducing more seams, while decreasing $\closeweight$ can reduce the seams but allow for a larger approximation error.
    \item Increasing $\barrierweight$ induces a stronger penalty for $\gamma$ values close to the boundary of its feasible range, which can limit the local variation of the resulting rulings and lead to higher approximation errors. Thus, we typically set $\barrierweight$ to a small value close to zero. 
    \item Since the Laplacian term promotes the smoothness of deformation and indirectly penalizes the deformation magnitude, increasing $\lapweight$ can lead to lower deviation from the reference surface but with more seams, while decreasing $\lapweight$ can allow for more deviation from the reference surface with fewer seams. 
    \item A larger $\edgelengthweight$ induces more restrictions on the edge length change, which can increase approximation errors but reduce the seams; on the other hand, a smaller $\edgelengthweight$ allows for more changes of edge length to adapt to the reference shape, thus decreasing the approximation error while increasing the seam length.
\end{itemize}

\begin{figure}[t]
    \centering
    \includegraphics[width=\linewidth]{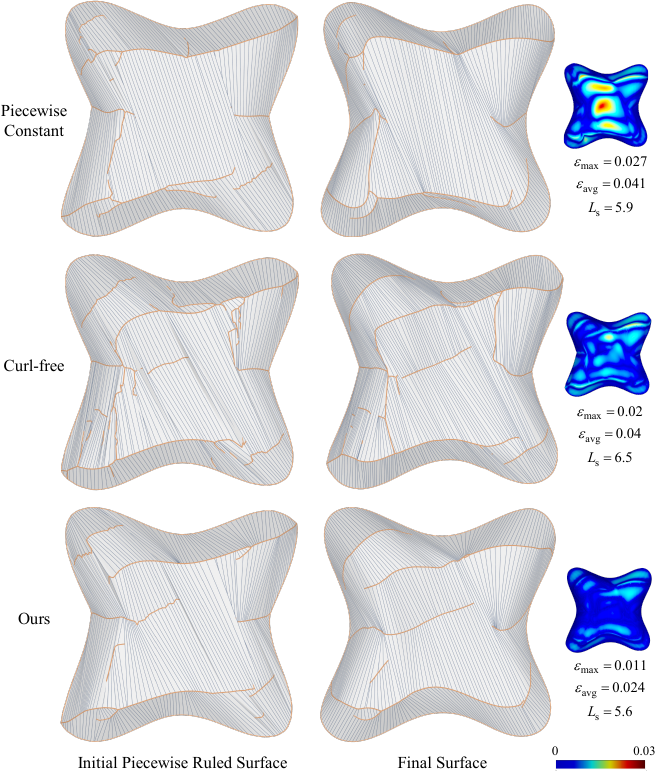}
    \caption{For the reference surface in Fig.~\ref{fig:pipeline}, our ruling vector field model leads to a result with lower approximation errors than alternative approaches based on piecewise constant ruling direction fields (top: simple replacement using piecewise constant ruling directions for optimization; middle: additionally enforcing the curl-free condition from~\cite{Vekhter2019}).}
    \label{fig:test-replace}
\end{figure}

\paragraph{Effectiveness of ruling direction field model}
Our model of non-constant ruling direction on a face plays an important role in our optimization. This is demonstrated in Fig.~\ref{fig:test-replace}, where we compare our method with alternative approaches based on the conventional piecewise constant ruling directions, using the same reference surface as in Fig.~\ref{fig:pipeline}. In the top row, we simply represent the ruling directions as a piecewise constant vector field while keeping all other components of our pipeline unchanged. In this case, the term $\edgegeoderrfunc$ in Eq.~\eqref{eq:Geod} requires the ruling directions on two adjacent faces to be parallel when unfolded to a common plane. Consequently, the joint optimization in Eq.~\eqref{eq:JointOptTargetFunction} results in an initial surface that is approximately piecewise cylindrical with almost parallel rulings in each patch, leading to higher approximation errors and longer seams in the final surface. In the middle row of Fig.~\ref{fig:test-replace}, we further replace the target term $\edgegeoderrfunc$ in Eq.~\ref{eq:CombinedErr} with a term $\edgecurlerrfunc$ enforcing the curl-free condition for geodesic vector fields proposed in~\cite{Vekhter2019}: $\edgecurlerrfunc(e_{ij}) = \|\overline{\mathbf{s}}_i \cdot \overline{\mathbf{e}}_{ij} - \overline{\mathbf{s}}_j \cdot \overline{\mathbf{e}}_{ij}\|^2$, where $\overline{\mathbf{e}}_{ij}$ is the unit edge vector for $e_{ij}$, and $\overline{\mathbf{s}}_i, \overline{\mathbf{s}}_j$ are normalized ruling directions on the two faces adjacent to $e_{ij}$. The resulting surface aligns better with the reference shape, but still performs worse than our method in both the approximation accuracy and the seam length. This example shows that our ruling direction field model can better capture the local variation of the rulings and enable more effective optimization.

\begin{figure}[t]
    \centering
    \includegraphics[width=\linewidth]{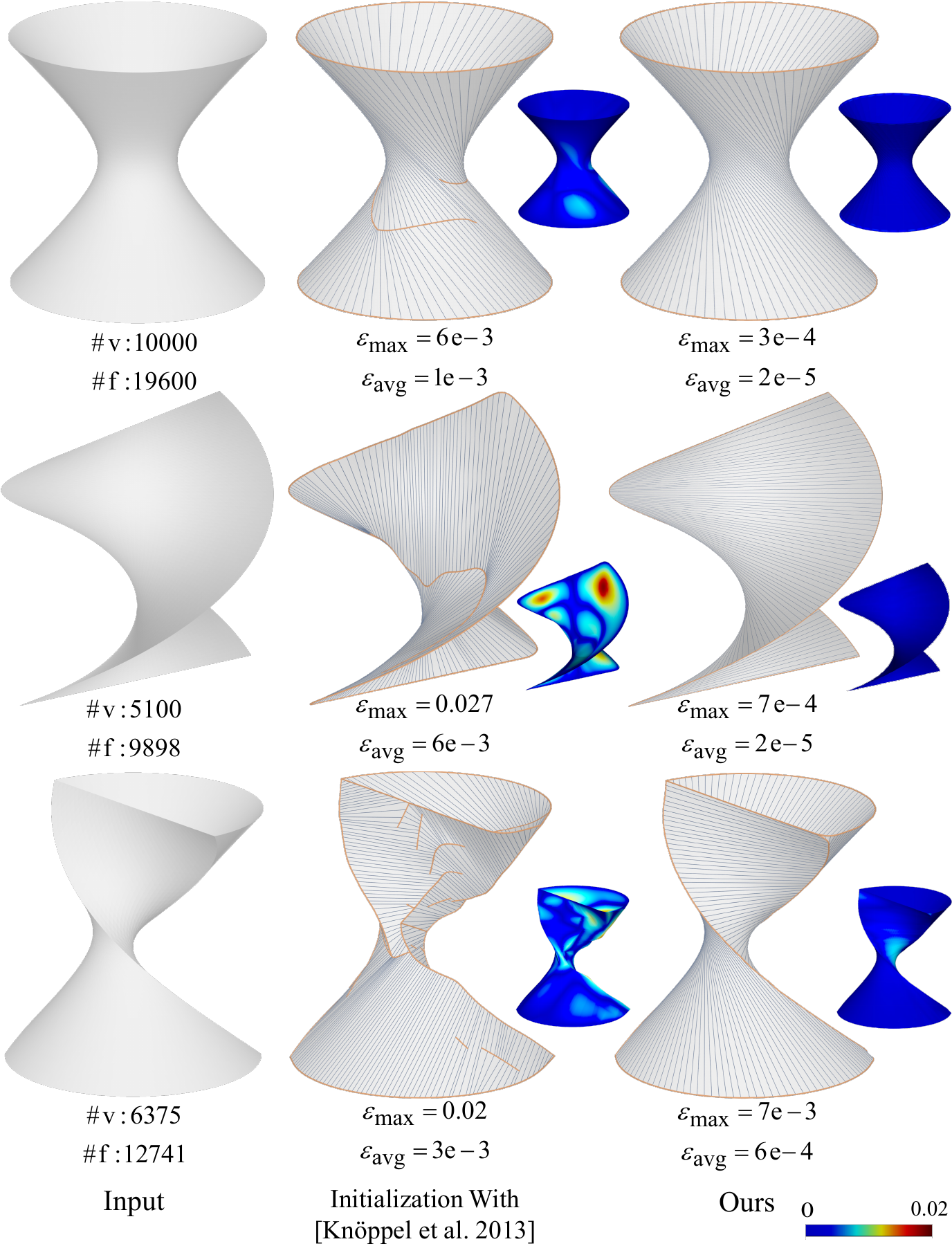}
    \caption{Thanks to our initialization strategy, our method can recover (piecewise) ruled target surfaces (top: a one-sheeted paraboloid; middle: a helicoid; bottom: a piecewise ruled surface consisting of two pieces each from one of the previous two surfaces), with rulings and seams well aligned with the ground truth. In comparison, initialization using the smooth vector field from~\cite{knoppel2013globally} leads to unnecessary seams and higher approximation errors.}
    \label{fig:typical}
\end{figure}

\paragraph{Effectiveness of initialization}
Our vector field initialization strategy presented in Sec.~\ref{sec:InitVectorField} is important for producing piecewise ruled surfaces with good approximation accuracy and without unnecessary seams. This is demonstrated in Fig.~\ref{fig:typical}, where we apply our method to three special target shapes: a one-sheeted paraboloid and a helicoid (both of which are classical ruled surfaces), and a piecewise ruled surface consisting of two patches from the two aforementioned surfaces respectively. Our method successfully recovers all three surfaces with low approximation errors, with both the rulings and seams well aligned with the ground truth. This is because our initialization already produces a vector field aligned with the ground-truth rulings, and the subsequent optimization quickly converges to the target shapes. For comparison, we also use the method from~\cite{knoppel2013globally}  to initialize a globally smooth vector field. The final results not only have higher approximation errors, but also introduce additional seams that are not present in the target shapes.  
This example verifies the necessity of our initialization strategy for effective optimization.

\paragraph{Computational efficiency} 
Finally, Table~\ref{tab:runtime} compares the computational time required by each method for the results shown in Fig.~\ref{fig:compare} and Fig.~\ref{fig:compare1}. We can see that our method runs faster than all other methods on these examples. Further data for the results in Fig.~\ref{fig:nu-comp} are provided in Appendix~\ref{sec:runtime}. These data demonstrate the efficiency of our computational approach.

%% file: conclusion.tex
\section{Limitations and Future Work}
Currently, our method only considers geometric approximation of the target and does not enforce other domain-specific conditions such as structural stability and accessibility for tools. Incorporating such conditions will be beneficial for practical applications.

In addition, the final piecewise ruled surface produced by our method may contain self-intersections. For applications that only require a discrete set of rulings (e.g., the string art in Fig.~\ref{fig:RuledSurfaceExamples}), such self-intersections may be acceptable since the gap between adjacent rulings can accommodate the penetrating rulings. Otherwise, we could potentially adopt the approach of~\cite{Yu2021} to prevent self-intersections, which will be an interesting future work.

Furthermore, our piecewise construction always produces sharp edges at the seams. This is necessary for target surface areas with positive Gaussian curvature: as the interior of a ruled surface patch can only have non-positive Gaussian curvature, the sharp edges are needed for `concentrating' the positive Gaussian curvature to the patch boundaries. However, for surface areas of non-positive Gaussian, it is possible to approximate them using ruled surface patches with continuous tangent planes but discontinuous ruling directions at the seams, producing a smooth appearance~\cite{flory2013ruled}. An interesting future work would be to adaptively enforce such tangent continuity conditions according to local curvature.

\begin{table}[t]
\caption{The computational time (in minutes) required by different methods for the models in Fig.~\ref{fig:compare} and Fig.~\ref{fig:compare1}, and the number of vertices and faces in the input models.}
\label{tab:runtime}
\centering
\setlength{\tabcolsep}{3pt}
%\resizebox{\linewidth}{!}{ 
\begin{tabular}{c|ccc|ccc}
\hline
\multirow{2}{*}{Model} & \multicolumn{3}{c|}{Fig.~\ref{fig:compare}} & \multicolumn{3}{c}{Fig.~\ref{fig:compare1}} \\
                       & Lilium   & Snail   & Bunny  & Bob   & Airport  & Kitten  \\ \hline
\#v                    & 3389     & 1000    & 3301   & 3087  & 3691     & 12414   \\
\#f                    & 6590     & 1909    & 6589   & 6174  & 7214     & 24828   \\
\cite{Stein2018}                   & 11.5     & 2.3     & 12.3   & 10.7  & 3.5      & 55.7    \\
\cite{Zhao2022}                   & 15.6     & 1.5     & 13.4   & 8.5   & 10.8     & 43.6    \\
\cite{Zhao2023}                   & 6.8      & 3.1     & 6.0    & 4.4   & 4.9      & 36.2    \\
Ours                   & 5.3      & 1.2     & 5.4    & 4.3   & 2.7      & 24.8    \\ \hline
\end{tabular}
%}
\end{table}

%% file: supp.tex
\section{Derivation of Eq.~(\ref{eq:PointRuilngVector})}
\label{sec:RulingDiretionDerivation}
From Eq.~\eqref{eq:LocalFaceModel}, we have
\[
{\mathbf{q}}_f(u, v) = \mathbf{o}_f + v  \facerule_f  +  (u + \gamma_f u v )\mathbf{c}_f.
\]
Therefore, if we $\mathbf{o}_f$ as the origin and $(\facerule_f, \mathbf{c}_f)$ as the axes of the local coordinate, then the local coordinates $(x_{\mathbf{s}},y_{\mathbf{s}})$ of a point with parameters $(u, v)$ satisfies
\begin{align*}
	&( v, u + \gamma_f u v )  = (x_{\mathbf{s}}, y_{\mathbf{s}})\\
	\Rightarrow~&v = x_{\mathbf{s}},~ u + \gamma_f u v = y_{\mathbf{s}}\\
	\Rightarrow~& u + \gamma_f u x_{\mathbf{s}} = y_{\mathbf{s}}\\
	\Rightarrow~& u = \frac{y_{\mathbf{s}}}{1 + \gamma_f x_{\mathbf{s}}}
\end{align*}
For the value $\frac{y_{\mathbf{s}}}{1 + \gamma_f x_{\mathbf{s}}}$ to be well-defined, we must have $1 + \gamma_f x_{\mathbf{s}} \neq 0$.
Then, using Eqs.~\eqref{eq:TaylorApproximation} and \eqref{eq:LinearDependence}, the ruling direction at $\mathbf{s}$ has the form
\[
 \mathbf{r}_f + u \gamma_f \mathbf{c}_f
 = 
 \mathbf{r}_f +  \frac{y_{\mathbf{s}}}{1 + \gamma_f x_{\mathbf{s}}}\gamma_f \mathbf{c}_f.
\]
Scaling this vector by $1 + \gamma_f x_{\mathbf{s}}$ (which changes its length but does not change the direction as $1 + \gamma_f x_{\mathbf{s}} \neq 0$), we obtain the ruling direction
\[
\facepointrule_f({\facept}) =  (1 + \gamma_f x_{\mathbf{s}}) \facerule_f  + \gamma_f y_{\mathbf{s}} \mathbf{c}_f.
\]

\section{Proof that Eq.~(\ref{eq:gammaCondition}) Implies $1 + \gamma_f x_{\mathbf{s}}  \neq 0$}
\label{sec:RulingConditionProof}
\begin{proof}
Since
$F(x_{\mathbf{s}}) = 1 + \gamma_f x_{\mathbf{s}}$ is an affine function of $x_{\mathbf{s}}$, it attains extremum values on a face at the points with the extremum values of $x_{\mathbf{s}}$, which must be the face vertices. If Eq.~(\ref{eq:gammaCondition}) is satisfied, then both the minimum and maximum values of $F(x_{\mathbf{s}})$ on the face are positive. Hence, $F(x_{\mathbf{s}}) \neq 0$ on the whole face.
\end{proof}

\section{Parameter Settings}
\label{eq:ParamSetting}
In vector field initialization, we usually set the start and end weights for $\unitweight$ as 1.0 and 10.0, 0.1 and 10.0 for $\npfacealignweight$. $\pfacealignweight$ always maintains a relatively small value, 0.01.

During the joint optimization, $\nu$ is an important parameter that affects the number of generated patch boundaries. We set $\nu_{\min}$ to the maximum values of $\combedgeerr$ across all edges for the initial mesh shape and ruling directions, while $\nu_{\min}$ is typically set to a value in $[0.05,0.06]$ in our experiments. For most models such as Bear, Bird, Fertility, and Kitten, we set $\nu_{\min}$ as 0.05. For Eight, Squirrel and Bob, we set $\nu_{\min}$ as 0.06. Increasing $\nu_{\min}$ will filter out some details of the model while reducing patch boundaries: for example, we set it to 0.07 for Bunny in Fig~\ref{fig:teaser}, 0.08 for Bust, and 0.09 for Rocket and Lilium. We need to relax the regularization weight appropriately to allow for this filtering. This is what we did in Fig.~\ref{fig:nu-comp}. On the other hand, reducing $\nu_{\min}$ will better approximate the target shape: for this purpose, we set it to 0.04 for Snale and Airport.

\section{Numerical Solver for Problem~(\ref{eq:InitOpt})}
\label{sec:MMSolver}
In the following, we use $\mathbf{Y}$ to denote a vector that concatenates all variables $\{\mathbf{y}_f\}$. We solve the optimization problem using a majorization-minimization (MM) solver~\cite{lange2016mm}. The key idea is that in each iteration, we construct a convex quadratic surrogate function $\overline{H}(\mathbf{Y}\mid\mathbf{Y}^{(k)})$ for the target function $H(\mathbf{Y})$ according to the current variable values $\mathbf{Y}^{(k)}$. This surrogate function should satisfy
\begin{equation}
    \begin{aligned}
        \overline{H}(\mathbf{Y}\mid\mathbf{Y}^{(k)}) \geq 
        H(\mathbf{Y})~~\forall~\mathbf{Y},\quad\text{and}~~~~
        \overline{H}(\mathbf{Y}^{(k)}\mid\mathbf{Y}^{(k)}) = {H}(\mathbf{Y}^{(k)}).
    \end{aligned}
    \label{eq:MMProperty}
\end{equation}
That is, $\overline{H}(\mathbf{Y}\mid\mathbf{Y}^{(k)})$ bounds $H(\mathbf{Y})$ from above, and has the same value as $H(\mathbf{Y})$ at $\mathbf{Y}^{(k)}$.
We then use $\overline{H}(\mathbf{Y}\mid\mathbf{Y}^{(k)})$ as a proxy for the target function, and minimize it to obtain the updated variable variables $\mathbf{Y}^{(k+1)}$, i.e.,
\[
    \mathbf{Y}^{(k+1)} = \argmin\nolimits_{\mathbf{Y}} ~\overline{H}(\mathbf{Y}\mid\mathbf{Y}^{(k)}).
\]
This process is repeated until the solver converges. Due to the properties~\eqref{eq:MMProperty}, the MM solver is guaranteed to decrease the target function $H$ in each iteration until convergence. 

To derive the surrogate function $\overline{H}$, we note that $\smoothterm$ is already convex and quadratic.
Additionally, on faces with non-negative Gaussian curvature (i.e., any face $f$ where $|\asympdirset_f|$ = 1), the term $\initalignterm$ is also convex and quadratic. For these terms, their surrogate functions are the same as themselves.
For each term $\initalignterm$ on a face with negative Gaussian curvature and the term $\unitterm$, we use the following surrogate functions respectively:
\[
\surroginitalignterm(\mathbf{y}_f \mid \mathbf{y}_f^{(k)})
 = \|\mathbf{y}_f \cdot \mathbf{t}_{\ast}^{(k)} \|^2, \quad
\surrogunitterm(\mathbf{Y} \mid \mathbf{Y}^{(k)})  = \sum_{f \in \mathcal{F}} \left\|\initlocalcoord_f - \frac{\initlocalcoord_f^{(k)}}{\left\|\initlocalcoord_f^{(k)}\right\|}\right\|^2,
\]
where $\mathbf{t}_{\ast}^{(k)} = \argmin_{\mathbf{t} \in \asympdirset_f} \| \mathbf{y}_f^{(t)} \cdot \mathbf{t}\|^2$.
Then we derive a convex quadratic surrogate function in the following form:
\begin{align*}
\overline{H}(\mathbf{Y}\mid\mathbf{Y}^{(k)})  = ~& \smoothterm + \unitweight \surrogunitterm(\mathbf{Y} \mid \mathbf{Y}^{(k)}) 
+ 
\npfacealignweight \sum\nolimits_{f \in \mathcal{F}_{\textrm{np}}} \surroginitalignterm(\mathbf{y}_f \mid \mathbf{y}_f^{(k)})\\
& + \pfacealignweight \sum\nolimits_{f' \in \mathcal{F}_{\textrm{p}}} \initalignterm(f').
\end{align*}
We then update $\mathbf{Y}$ via 
\[
    \mathbf{Y}^{(k+1)} = \argmin\nolimits_{\mathbf{Y}} \overline{H}(\mathbf{Y}\mid\mathbf{Y}^{(k)}),
\]
which amounts to solving a linear system with a sparse symmetric positive definite matrix.

For initialization of the MM solver, we construct a proxy function 
\[
\widehat{H} = \mathbf{Y}^T \mathbf{M}_{H} \mathbf{Y}
\]
that approximates the target function in Eq.~\ref{eq:InitOpt}, where $\mathbf{M}_{H}$ is a sparse symmetric positive definite matrix. We compute the eigenvector $\widehat{\mathbf{Y}}$ of $\mathbf{M}_{H}$ corresponding to its smallest eigenvalue, which is the minimizer of $\widehat{H}$ among all unit vectors. We then normalize the components of $\widehat{\mathbf{Y}}$ for each face to obtain the initial value ${\mathbf{Y}}^{(0)}$ for the MM solver.
To construct $\widehat{H}$, we discard the unit-length terms, and replace each $\initalignterm$ on a face $f$ of negative Gaussian curvature with
\[
   \proxyinitalignterm(\mathbf{y}_f) = \| \mathbf{y}_f \cdot \mathbf{e}_f \|^2, 
\]
where $\mathbf{e}_f$ is a unit vector parallel to the principal direction of $f$ with the largest magnitude of principal curvature. In this way, $\proxyinitalignterm(\mathbf{y}_f)$ requires $\mathbf{y}_f$ to align with the other principal direction, which is the solution to the following proxy problem that minimizes the total alignment error with all asymptotic directions on the face:   
\[
    \min_{\|\mathbf{v}\| = 1}  \sum\nolimits_{\mathbf{t} \in \asympdirset_f} \| \initlocalcoord_f \cdot \mathbf{t}\|^2.
\]
We then derived our proxy function as:
\[
\widehat{H} = \smoothterm + \npfacealignweight \sum\nolimits_{f \in \mathcal{F}_{\textrm{np}}} \| \mathbf{y}_f \cdot \mathbf{e}_f \|^2\\
+ \pfacealignweight \sum\nolimits_{f' \in \mathcal{F}_{\textrm{p}}} \initalignterm(f').
\]

We initialize $\unitweight$ and $\npfacealignweight$ to relatively small values $\unitweight^{\min}$ and $\npfacealignweight^{\min}$ so that the optimization can focus on the smoothness conditions initially, then gradually increase them to large values $\unitweight^{\max}$ and $\npfacealignweight^{\max}$ to better enforce the unit-length conditions and the alignment with asymptotic directions. In our experiments, we set $\unitweight^{\min} = 1$, $\npfacealignweight^{\min} = 0.1$, $\unitweight^{\max} = 10$, $\npfacealignweight^{\max} = 10$, and $\pfacealignweight = 0.01$.

\begin{figure}[t]
	\includegraphics[width=\linewidth]{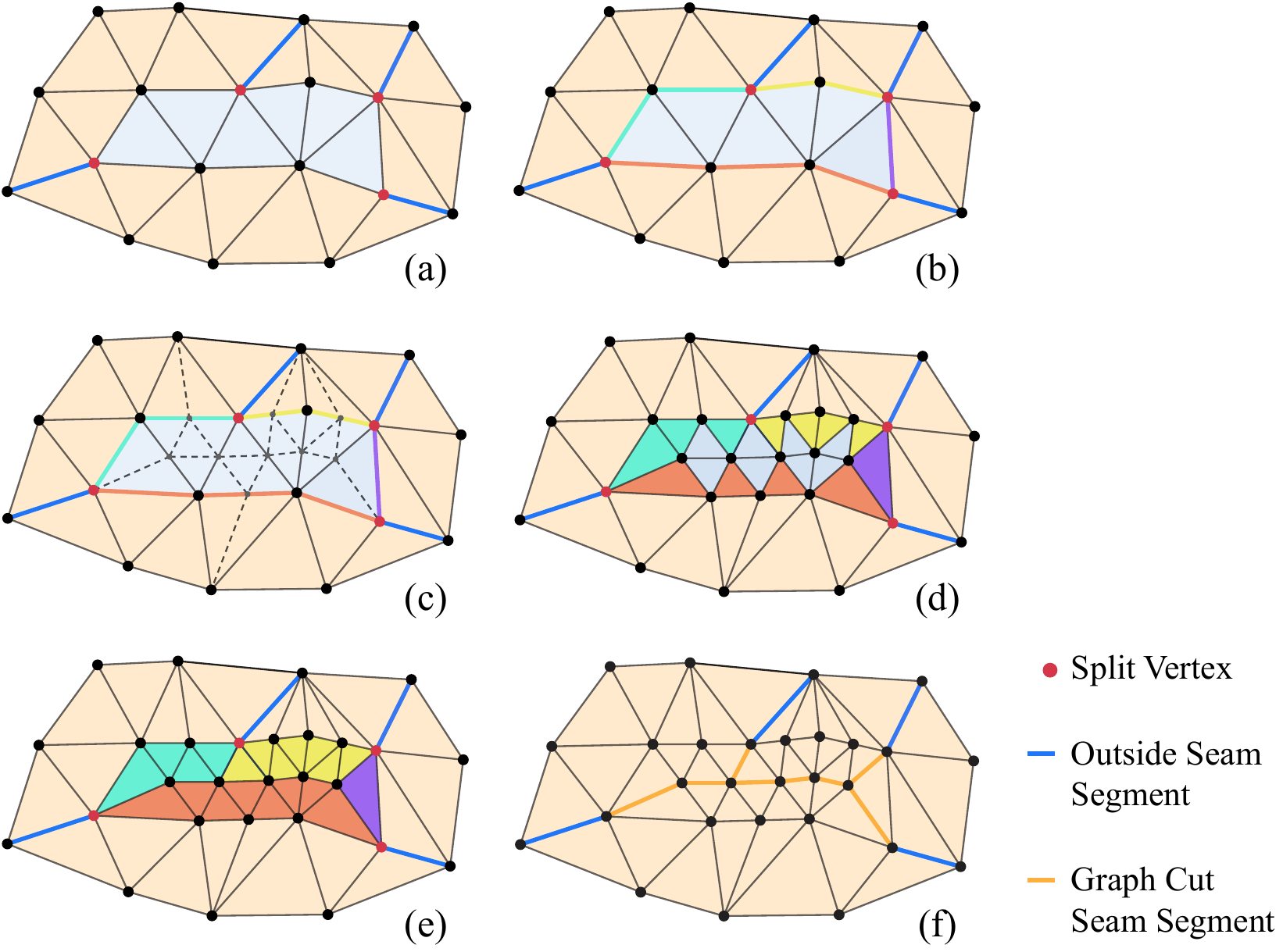}
	\caption{Illustration of our graph cut optimization to derive seam segments inside a connected region (shown in grey in (a)).}
	\label{fig:GraphCut}
\end{figure}

\section{Graph Cut Optimization}
\label{sec:GraphCut}
If a connected region $\facecomponent$ contains more than one face, we perform graph cut optimization to segment the interior of $\facecomponent$ based on the split vertices at the boundary of $\facecomponent$ and use the cut edges as the ruled surface patch boundary segments inside $\facecomponent$. 
The process is illustrated in Fig.~\ref{fig:GraphCut}, where Fig.~\ref{fig:GraphCut}~(a) shows the region $\facecomponent$ in grey.
We assume there are $N$ split vertices on the boundary of $\facecomponent$ ($N \geq 2$), dividing the boundary into $N$ segments. We assign a label $i$ ($i=1,\ldots,N$) to each segment (shown in different colors in Fig.~\ref{fig:GraphCut}~(b)).
We then subdivide each face inside $\facecomponent$ according to the number of different labels on its edges (see Fig.~\ref{fig:GraphCut}~(c)): If there are two labels (i.e., two of its edges are on the boundary of $\facecomponent$ and assigned with different labels), then we subdivide the face into two sub-faces by splitting the non-boundary edge at its mid-point. Otherwise, we subdivide the face into four sub-faces by splitting each of its edges along their mid-points.
For faces outside $\facecomponent$, if it contains an edge that is split due to the above subdivision, we also subivide it into two sub-faces along the split edge to ensure all faces are triangles.
We denote the set of all sub-faces in $\facecomponent$ as $\overline{\facecomponent}$.
We then derive a dual graph for the sub-faces in $\overline{\facecomponent}$ and perform a graph cut on this dual graph as follows.
First, we note that after the subdivision, each sub-face at the boundary of $\facecomponent$ is attached to exactly one boundary edge of $\facecomponent$. We assign to such a sub-face the same label $i$ as the boundary edge it is attached to (shown in Fig.~\ref{fig:GraphCut}~(d) using the same colors as the corresponding boundary edge colors in Fig.~\ref{fig:GraphCut}~(c)).
We then solve a graph cut optimization problem to label the remaining sub-faces in $\overline{\facecomponent}$ (see Fig.~\ref{fig:GraphCut}~(e)). Specifically, for each remaining sub-face $f$ we have a label variable $L_f \in \{1, \ldots, N\}$. We determine the labels via an optimization:
\begin{equation}
    \begin{aligned}
        \min_{\{L_f\}} ~~\sum\nolimits_{e \in \facecompintedgeset} \edgecutcost(e)  
        + \facelabelcostweight \sum\nolimits_{f \in \unlabelledfaceset} \facelabelcost(f),
    \end{aligned}
\end{equation}
where $\unlabelledfaceset$ denotes the set of sub-faces that require labeling, and $\facecompintedgeset$ denotes the set of interior edges of $\overline{\facecomponent}$ with at least one adjacent sub-face requiring labelling. $\edgecutcost$ is the cost function for an edge $e$ depending on the label of its two adjacent sub-faces: if both sub-faces have the same label, then $\edgecutcost(e) = 0$; otherwise, $\edgecutcost(e)$ equals the length of $e$. $\facelabelcost$ is a label cost function for a sub-face $f$ and equals the shortest geodesic distance from $f$ to a boundary sub-face with the same label as $f$. This optimization produces a segmentation consistent with the boundary sub-face labels, while reducing the total length of the cut edges, simplifying the shape of boundaries between areas of different labels. We solve this optimization problem using the \textsc{GCoptimization} library\footnote{\url{https://vision.cs.uwaterloo.ca/files/gco-v3.0.zip}}. After the graph-cut optimization, the edges between faces of different labels are used as seam segments inside $\facecomponent$ (see Fig.~\ref{fig:GraphCut}~(f)).

\section{More Data on Computational Efficiency}
\label{sec:runtime}
Table~\ref{tab:gallery-runtime-transposed} shows our method's computational time for the examples in Fig.~\ref{fig:gallery}.
\begin{table}[t]
\caption{The runtime of our method (in minutes) for the models in Fig.~\ref{fig:gallery}, and the number of vertices and faces in these models.}
\label{tab:gallery-runtime-transposed}
\centering
\setlength{\tabcolsep}{2pt}
\resizebox{\linewidth}{!}{ 
\begin{tabular}{lcccccccc}
\toprule
 & Bust & Fertility & Eight & Bird & Bear & Rocket & Snale & Squirrel \\
\midrule
\#v   & 11426 & 10463 & 8030 & 5710 & 5313 & 8152 & 2705 & 6298 \\
\#f   & 22848 & 20938 & 16064 & 11416 & 10622 & 16300 & 5349 & 12592 \\
Time (min) & 15.4 & 34.6 & 12.8 & 7.2 & 6.7 & 9.3 & 3.5 & 12.3 \\
\bottomrule
\end{tabular}
}
\end{table}